\documentclass[a4paper, 11pt]{article}
\usepackage{amsmath, amsthm,amssymb,mathtools}
\usepackage{tgtermes}
\usepackage{enumitem}
\usepackage[T1]{fontenc}
\usepackage{framed}

\usepackage{times}
\usepackage{graphicx}
\usepackage{xcolor}
\usepackage{subcaption}

\newtheorem{thm}{Theorem}
\newtheorem{lem}{Lemma}
\newtheorem{prop}{Proposition}
\newtheorem{cor}{Corollary}

\newtheorem{rmk}{Remark}
\newtheorem{defin}{Definition}
\newtheorem{ex}{Example}
\newtheorem{pb}{Problem}

\newcommand{\bp}{\begin{pb}\rm}
\newcommand{\ep}{\end{pb}}
\newcommand{\br}{\begin{rmk}\rm}
\newcommand{\er}{\end{rmk}}
\newcommand{\bdefin}{\begin{defin}\rm}
\newcommand{\edefin}{\end{defin} }
\newcommand{\bex}{\begin{ex}\rm}
\newcommand{\eex}{\end{ex}}
\newcommand{\lcage}{\langle}
\newcommand{\rcage}{\rangle}

\newcommand{\newc}{C}
\newcommand{\newa}{D}

\newcommand{\bthm}{\begin{thm}}
\newcommand{\ethm}{\end{thm}}
\newcommand{\blem}{\begin{lem}}
\newcommand{\elem}{\end{lem}}
\newcommand{\bprop}{\begin{prop}}
\newcommand{\eprop}{\end{prop}}
\newcommand{\bcor}{\begin{cor}}
\newcommand{\ecor}{\end{cor}}

\usepackage{algorithm}
\usepackage{algorithmic}

\usepackage{xargs}                      

\setlist[itemize]{noitemsep, nolistsep}
\setlist[enumerate]{noitemsep, nolistsep}

\begin{document}
\vspace*{-1cm}
\begin{center}
{\large \sc On the complexity of recognizing Stick, BipHook and Max Point-Tolerance graphs }
\bigskip

Irena Rusu\footnote{Irena.Rusu@univ-nantes.fr}

{\it Nantes Universit\'e, \'{E}cole Centrale Nantes,\\ CNRS, LS2N, UMR 6004, F-44000 Nantes, France
}
\end{center}
\bigskip\bigskip

\begin{center}
\begin{minipage}[h]{13cm}
\paragraph{Abstract}
 {\small Stick graphs are defined as follows. Let $A$ (respectively $B$) be a set of vertical (respectively horizontal) segments in the plane such that 
 the bottom endpoints of the segments in $A$ and the left endpoints of the segments in $B$ lie on the same ground straight line with slope $-1$. The Stick graph
defined by $A$ and $B$, which is necessarily bipartite, is the intersection graph of the segments in $A$ with the segments in $B$.

  We answer an open problem by showing that recognizing Stick graphs is NP-complete. This result allows us to easily solve two other open problems, namely the recognition of BipHook graphs and of max point-tolerance graphs. We show that both of them are NP-complete problems. 
}
 
 \end{minipage}
 \end{center}
 \bigskip
 \bigskip
 
\section{Introduction}

Intersection graphs represent the intersections between the objects in a given collection, and are extremely diverse. Many geometric objects in one or two dimensions have been used to define classes of intersection graphs,  like segments, arcs on a cycle, trapezoids, curves etc. The books \cite{brandstadt1999graph,mckee1999topics} present many results on this topic. The considerable amount of work on intersection graphs, especially when the objects are geometric, is due both to their applications, for instance to electrical networks, nano PLA-design, computational biology, traffic control, statistics, psychology \cite{sinden1966topology,mckee1999topics,shrestha2011two,halldorsson2011clark,baruah2013intersection}, and to the variety of problems they raise.

Given a collection $\mathcal{C}$ of objects, the intersection graph of $\mathcal{C}$ is the graph with vertex set $\mathcal{C}$ whose edges are
the pairs $(c_1,c_2)\in \mathcal{C}\times \mathcal{C}$ such that $c_1\neq c_2$ and $c_1\cap c_2\neq\emptyset$.  
When $\mathcal{C}$ is a class of segments on a straight line, the resulting intersection graph is called an {\em interval graph}. Interval graphs have been  intensively studied, and many classical problems are
linear when restricted to them: recognition \cite{booth1976testing}, independent set \cite{gavril1974intersection,rose1976algorithmic}, isomorphism testing \cite{booth1976testing} etc. (see \cite{brandstadt1999graph} for more examples), although a few of them are NP-complete (for instance, maximum cut \cite{adhikary2021complexity}). The generalization to straight line segments in a plane defines the class of SEG graphs, for which -- at
the opposite of interval graphs -- all the classical problems whose algorithmic complexity is known are NP-complete or worse. In particular, recognizing SEG graphs is NP-hard \cite{kratochvil1994intersection} and $\exists\mathbb{R}$-complete \cite{schaefer2009complexity}. This shows that, even when restricted to very simple geometric objects such as segments, intersection graphs cover a wide variety of structures, and thus offer an impressive diversity of problems that are both difficult and interesting. 

In this paper we focus on classes of intersection graphs of straight segments on a grid. A  {\em grid intersection graph} (defined in \cite{hartman1991grid} and abbreviated GIG) is the intersection graph of $A\cup B$, where $A$ (respectively $B$) is a set of vertical 
(respectively horizontal) segments in the plane such that two vertical (respectively horizontal) segments are disjoint. 
Grid intersection graphs are thus bipartite graphs. (Note that sometimes the same terminology of grid intersection graphs is
used for the larger class where the parallel segments are not required to be disjoint.) Despite their apparent simplicity, recognizing grid intersection graphs is NP-complete \cite{kratochvil1994special}. In  \cite{chaplick2018grid}, the
authors investigate grid intersection graphs by defining several subclasses and studying the relations between them as well as their interval and order dimension. Among them, the class of Stick graphs is a minimal one for which the complexity of the recognition problem is not known. Larger
than the class of Stick graphs, but very closely related to it, is the class of bipartite hook graphs, for which the same question is  open too.

{\em Stick graphs}  are the grid intersection graphs whose vertical (respectively horizontal) segments have their bottom (respectively left) endpoint, also called the {\em origin} of the segment, on the same ground straight line with slope -1. A {\em hook} is a couple made of a vertical segment and an horizontal segment whose bottom
and respectively left endpoint coincide. The common endpoint is the {\em center} of the hook. {\em Bipartite hook graphs} (or BipHook graphs) are the {\em bipartite} 
intersection graphs of a set of hooks whose centers lie on a ground straight line with slope -1.  It is shown in \cite{chaplick2018grid} that Stick graphs
are strictly contained in BipHook graphs. Another generalization of Stick graphs is obtained by replacing the vertical and horizontal segments
with {\em grounded segments}, that are by definition the segments with an endpoint on the ground line and whose all other points are
above the ground line. The recognition of intersection graphs of grounded segments is
$\exists\mathbb{R}$-complete \cite{cardinal2018}.

Whereas the complexity of testing whether a given graph $G=(A\cup B, E)$ is a Stick graph is still open, the variants of the problem 
where the order of the origins along the ground line is known  either for both sets $A$ and $B$, or only for one of them, may be solved in linear time. For the former variant,
a characterization theorem and a quadratic algorithm based on it have been proposed in \cite{luca2018recognition2}. A different approach,
using a line that sweeps the ordered origins from left to right along the ground line and progressively computes the solutions has 
been proposed in \cite{chaplick2019recognizing2}, and yielded a linear algorithm for the former variant, as well as a quadratic algorithm for the latter 
variant. A third approach \cite{rusu2021forced}  extended the ideas in \cite{luca2018recognition2} to the more  general case where only the origins in $A$ are ordered, and proposed a left-to-right search that is able to build in linear time the digraph recording all the solutions ({\em i.e.}, the orderings of the origins in $A\cup B$ for which a Stick representation exists).
Another linear algorithm for the former variant thus follows, which consists in applying the third approach under the assumption that
only the origins in $A$ are ordered, and then testing whether the order of the origins in $B$ is admitted by the resulting digraph.

Another variant of  Stick graphs has been proposed in \cite{chaplick2019recognizing2}. In this case, the length
of the segment representing each vertex is provided. Recognizing Stick graphs with fixed length segments is NP-complete in the three cases: when no order is provided on the set
of origins, when only the origins in $A$ (equivalently, in $B$) are ordered, and when the origins in both $A$ and $B$ are ordered. Further
subvariants where isolated vertices are not admitted are also NP-complete, except in the third case, where this subvariant 
is polynomially solvable \cite{chaplick2019recognizing2}.

In this paper, we show that recognizing Stick graphs is NP-complete, and give two easy, but significant, corollaries of this result. The first
of them is that recognizing BipHook graphs is NP-complete too. The second one immediately follows from the first one: recognizing max point-tolerance graphs (also known as {\em hook graphs}, {\em heterozygosity graphs} or {\em p-BOX(1) graphs}) \cite{catanzaro2017max,halldorsson2011clark,hixon2013hook,thraves2015p} is also NP-complete. Max point-tolerance graphs are the (not necessarily bipartite) intersection graphs of hooks with (distinct) centers on a line with slope -1. Several characterizations are known for this class of graphs that received a lot of attention since about a decade ago (see the papers cited above), but 
the recognition problem is still open.

The paper is organized as follows. Section \ref{sec:Intro} introduces the notations and the Stick graph recognition problem ({\sc StickRec}). 
In Section \ref{sect:reduction} we describe the construction that allows us to show in Section \ref{sect:mainth} that {\sc StickRec} is NP-complete.
The NP-completeness of recognizing BipHook and max point-tolerance graphs is proved in Section \ref{sect:biph}. Section \ref{sect:conclusion}
is the conclusion.

\section{Definitions and notation}\label{sec:Intro}

The graphs we use are undirected and simple. Given a graph $G$, its vertex set is denoted by $V(G)$ and its set of edges by $E(G)$.
The {\em neighborhood} of a vertex $x$ is the set of vertices adjacent to $x$, and is denoted by $N(x)$. If $H$ is an induced
subgraph of $G$, then we also use the notation $N_H(x)$ for $N(x)\cap V(H)$.

\begin{figure}[t!]
\centering
\hspace*{-3cm}\scalebox{0.5}{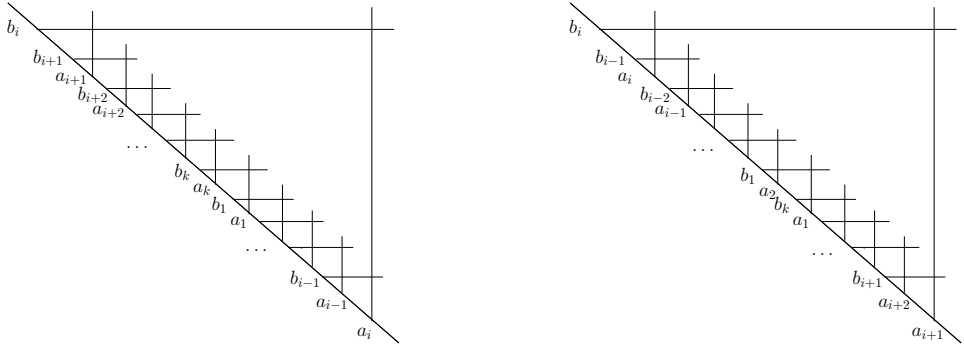}
\caption{\small The Stick representations of the cycle $a_1, b_1, a_2, b_2, \ldots, a_k, b_k$. The index $i$ is arbitrary, with $i\in\{1, 2, \ldots, k\}$.}
\label{fig:cycle}
\end{figure}

Given a Stick graph $G=(A\cup B,E)$, a {\em Stick representation} of $G$ consists in a set of vertical (respectively horizontal) segments,
one for each vertex in $A$ (respectively in $B$), whose origins lie on the ground line, and whose pairwise intersections correspond to
the edges in $E$. See Figure \ref{fig:cycle} for an example. 
We may assume, without loss of generality that the origins of these segments are distinct, so that they are totally ordered along
the ground line. The other endpoint of each segment is called its {\em tip}. 

For each vertex $c\in A\cup B$, its associated segment as well as the origin of the segment are also denoted by $c$.  
The interpretation of a notation is thus deduced from the context: the origins are {\em placed} or {\em positioned}, the segments {\em intersect}
and the vertices are {\em adjacent} with other vertices. The segments are called $A$-segments if they are vertical (correspond to vertices in $A$)
or $B$-segments if they are horizontal (correspond to vertices in $B$). The origins are also called $A$-origins or $B$-origins, according to the type of the segment. An origin $c_1$ is {\em before} another origin $c_2$ if $c_1$ is to the left of $c_2$ (equivalently, $c_1$ is higher than $c_2$) along the ground line. We denote this order by $c_1\prec c_2$. We call a {\em cage} the triangular region of the plan defined by the ground line, a $B$-segment $b$ and an $A$-segment $a$ such that $b\prec a$, and $a,b$ intersect.
The cage defined by $b$ and $a$ is denoted by $\lcage b, a\rcage$.

\br
The Stick representations of a $2k$-cycle $a_1, b_1, a_2, b_2, \ldots, a_k, b_k$,  are given in Figure~\ref{fig:cycle}. Each of them orders the $B$-origins along the ground line starting with a $B$-origin
$b_i$ (for some $i\in\{1, 2, \ldots, k\}$), followed by the other $B$-origins in increasing or decreasing circular order of the indices. Then we have either  $b_i\prec b_{i+1}\prec  \ldots \prec b_k\prec b_1\prec  \ldots \prec b_{i-1}$
or $b_i\prec b_{i-1}\prec \ldots\prec b_1\prec b_k\prec \ldots \prec b_{i+1}$ from left to right.  This is explained as follows.  There exists at least one pair $b_j, b_{j+1}$ of 
$B$-origins with consecutive indices (in a circular order), such that an origin $b_u$ with $u\neq j, j+1$  is placed between $b_j$ and $b_{j+1}$.
Then the interval delimited by $b_j$ and  $b_{j+1}$ contains all the other origins. Otherwise each path in the $2k$-cycle between an origin outside the interval and $b_u$ should contain an internal vertex whose corresponding
segment intersects $b_j, b_{j+1}$ or $a_{j+1}$, and this is false. We deduce that exactly one pair of $B$-origins  $b_j$ and $b_{j+1}$ exists with this property (the reciprocal inclusions are not possible), 
and all the other pairs of $B$-origins $b_h, b_{h+1}$ (in a circular order) are consecutive along the ground line. 
\label{rem:cycle}
\er

The problem we are interested in is defined as follows:
\bigskip

\noindent{\sc Stick Graph Recognition (StickRec)}

\noindent {\bf Input:} A bipartite graph $G=(A\cup B, E)$.

\noindent{\bf Output:} Does $G$ admit a Stick representation?
\bigskip

We will show that this problem is NP-complete. To this end, in Section \ref{sect:reduction} we describe our reduction and in Section \ref{sect:mainth} we prove the abovementioned result.

\section{The reduction}\label{sect:reduction}

Consider the following problem:
\newpage

\noindent{\sc Monotone 1-in-3SAT}

\noindent {\bf Input:} A set $\mathcal{K}$ of $m$ clauses with three literals each, such that each literal is a variable from the set $X=\{x_1, x_2, \ldots, x_n\}$.  

\noindent{\bf Output:} Is there a truth assignment for the variables in $X$ such that each clause contains exactly one true literal?
\medskip

The problem {\sc Monotone 1-in-3SAT} is NP-complete (see \cite{schaefer1978complexity} for 1-in-3SAT and problem L04 in \cite{Garey79} for the monotone variant).

\paragraph{Conventions}  We  make the assumption that in each clause the literals are distinct. In the contrary case, remark that the clause  $(x_i\vee x_i\vee x_j)$ has a truth assignment with exactly one true literal 
if and only if all the clauses $(x_i\vee x_j\vee u), (x_i\vee x_j\vee w), (x_j\vee u \vee w)$ satisfy the same property. There is only one such truth assignment, and it sets the  variable $x_j$ to true, and variables $x_i,u,w$ to false. Then each clause with two literals may be replaced by the abovementioned
set of three clauses. Moreover, each clause $(x_i\vee x_i\vee x_i)$ may be firstly replaced with the clauses $(x_i\vee x_i\vee a)$,
$(x_i\vee x_i\vee b),(x_i\vee a\vee b)$, where $a$ and $b$ are new variables, and then each of the two clauses with two identical literals may be  replaced again as before. We may therefore assume
that in each clause the literals are distinct.

Then the clause with literals $x_i, x_j, x_k$ where $i<j<k$ is uniquely written as $(x_i\vee x_j\vee x_k)$. We also assume that the clauses are totally ordered 
according to an arbitrary, but fixed, order and denoted by $K_1, K_2, \ldots, K_m$.

\begin{figure}[t!]
\centering
\vspace*{-2cm}
\hspace*{-3cm}\includegraphics[width=16cm,angle=270]{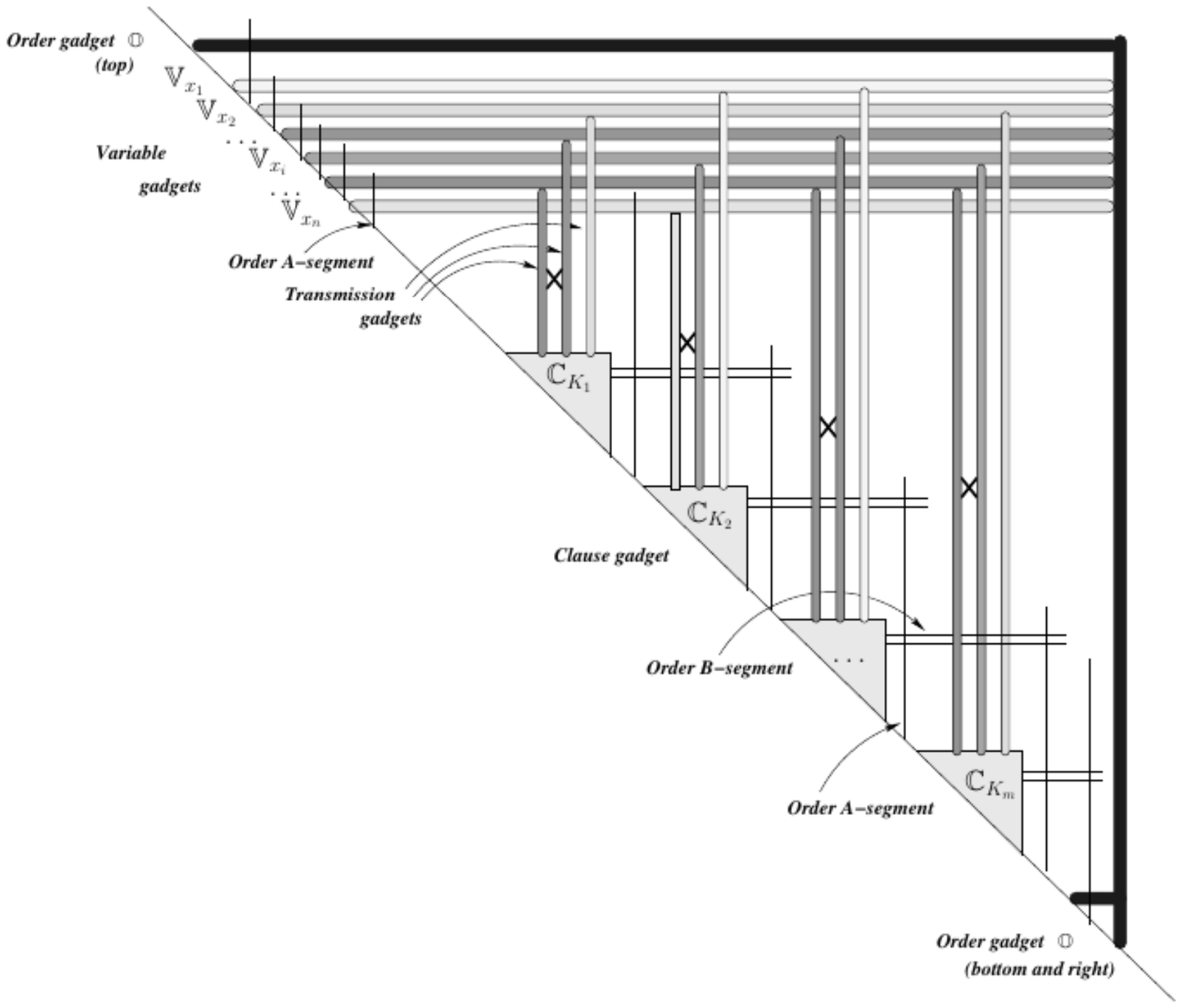}
\vspace*{-2cm}
\caption{\small  Overview of the construction. With $n=6$, the two first clauses are $K_1=(x_2\vee x_3\vee x_5)$
and $K_2=(x_1\vee x_4\vee x_6)$. Note here that the left-to-right order of the literals in the clause is the
right-to-left order of the transmission gadgets. The $\large \times$ sign between the transmission gadgets of the
second and third literal (the leftmost and the middle one, for each clause) indicate that these transmission gadgets are intertwined.}
\label{fig:overview}
\end{figure}

\bigskip

We build the instance $\mathbb{G}(X,\mathcal{K})$ of {\sc StickRec} associated with $X$ and $\mathcal{K}$ in the next subsections. The graph
$\mathbb{G}(X,\mathcal{K})$ contains variable, clause and transmission gadgets as well as an order gadget with numerous intersections, 
indicated in Figure~\ref{fig:overview} and explained in detail below. In the figure, a common point of two gadgets means that there exists at least one
intersection between two segments from the two gadgets. A clause gadget together with its three transmission gadgets is called a {\em clause-transmission gadget}.
Then, by definition, a {\em standard Stick representation} of $\mathbb{G}(X,\mathcal{K})$ is a representation that places the order, variable and clause-transmission gadgets 
as in Figure~\ref{fig:overview}. More precisely, it orders, from left to right, the variable gadgets $\mathbb{V}_i$ for $x_i$, $1\leq i\leq n$, in increasing order of $i$, 
and then the clause-transmission gadgets ($\mathbb{C}_{K_j}$ and the transmission gadgets rising from it) for $K_j$, $1\leq j\leq m$, in increasing order of $j$. 
This order is then used as follows. Each Stick representation of a clause gadget fixes a {\em state} (which is T or F, for true or false) for each literal, 
and outputs the states of the three literals towards the  variable gadgets, using the three transmission gadgets. Each variable gadget guarantees that 
the states communicated by the clauses are the same, and thus that a unique value T or F, identical to the common state, can be affected to the variable.

The intersections between the elements in Figure~\ref{fig:overview} are of three types. The intersections forcing the order of the gadgets in a standard Stick representation occur between
the order gadget, the order segments and one or two $B$-segments in each variable gadget. The intersections spreading the states of the literals occur
between a clause gadget and the three transmission gadgets corresponding to its three literals, as well as between each of the three transmission gadgets and 
the corresponding variable gadget. The intersections between gadgets, or between gadgets and order segments, not involved in the two tasks above are by-products of the first ones. We call them 
{\em incidental intersections}. The two former ones are finely defined according to their goals. The latter ones need only to note that:

\br
The fixed order of the variable gadgets and of the clause-transmission gadgets in the standard Stick representation allows us to  precisely identify all the incidental intersections.  
For instance, Figure~\ref{fig:overview} indicates that the variable gadget $\mathbb{V}_{x_i}$ of $x_i$ intersects the transmission gadgets of the literals $x_j$ with $j<i$ from all the clauses. Even more precisely, 
a $B$-segment from a  variable gadget whose tip is positioned, for instance, between two given clause-transmission gadgets intersects all the transmission gadgets 
of the  literals $x_j$ with $j<i$ from the clause gadgets preceding the tip. 
\label{rem:intersectionvartrans}
\er

In our construction, the incidental intersections are thus defined strictly according to the unique order of the gadgets used by all the standard Stick representations.   
Consequently, the existence of a Stick representation for the graph $\mathbb{G}(X,\mathcal{K})$  reduces to 
the existence of a standard Stick representation for the clause-transmission gadgets that output consistent values for the variable gadgets, which admit then Stick representations.

We now define the gadgets and the graph $\mathbb{G}(X,\mathcal{K})$. We use capital letters like $\mathbb{H}, \mathbb{C}$ for the graphs,
lowercase letters for the vertices/segments/origins with indices (indices mean  similarity, up to a point) and capital letters for the 
vertices/segments/origins with no symmetry. Exceptions to these rules are the  $A$-segments in the transmission gadgets, which
are indexed but use capital letters for a better readability.

Recall that a vertex, the segment representing it and the origin of the segment have the same notation. The interpretation of a notation is thus deduced from the context: the origins are {\em placed} or {\em positioned}, the segments {\em intersect}
and the vertices are {\em adjacent} with other vertices.

The descriptions of the gadgets and the construction of their Stick representations require many drawings, as well as numerous 
attempts to position the origins. In our drawings of the graphs, the white vertices are the vertices in $A$,
and the black ones are the vertices in $B$. In the Stick representations, when several origins are already positioned along the ground line, we say that a new origin 
has a {\em unique place} on the ground line if there is only one interval, either between two already positioned origins, or before the 
leftmost origin or after the rightmost origin, where the new origin can be placed such that its corresponding segment
intersects all the segments it has to intersect, and that are already drawn. When we say that an origin is  placed {\em immediately
before} or {\em immediately after} another origin $c$ we mean it is placed in the interval between the origin preceding $c$ and $c$,
respectively between $c$ and the origin following $c$.

\subsection{The handy gadget $\mathbb{H}$}

\begin{figure}[t!]
\centering
\hspace*{-3cm}\scalebox{0.5}{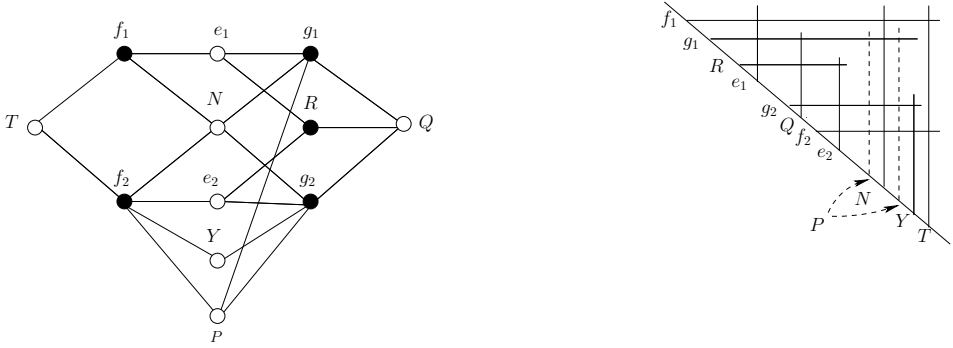}
\caption{\small The handy gadget (left) and its Stick representations (right). There are two possible positions for $P$ (dashed lines) and then $Y$
may be positioned on either side of $P$ and $N$ (three possible positions). With an aim of simplification, only one of the possible positions for $Y$ is drawn. }
\label{fig:gadgetO}
\end{figure}

The handy gadget $\mathbb{H}$ is used to constrain the relative positions of two or several segments. It is given in Figure~\ref{fig:gadgetO} (left). 

\bprop
Up to minor variants concerning the positions of $P$ and $Y$, the handy gadget $\mathbb{H}$ has a unique Stick representation, given in Figure~\ref{fig:gadgetO} (right).
\label{prop:stickRhandyG}
\eprop

\begin{proof}
Let us denote by $\mathbb{H'}$ the  graph obtained from the handy gadget by removing $P, Y$ and their incident edges. Note that $\mathbb{H'}$ is symmetric with 
respect to the (imaginary) horizontal line going through $T, N, R, Q$, and assume without loss of generality that we look for the Stick representations where $f_1\prec f_2$. The vertices $T, f_1, e_1, g_1, Q, g_2, e_2, f_2$ induce a
8-cycle of $\mathbb{H'}$ whose four $B$-vertices are $f_1, g_1, g_2, f_2$, in this order along the cycle. According to Remark \ref{rem:cycle} and to the assumption that
$f_1\prec f_2$, there exist four Stick representations for this cycle, given by the orders: $(i)$ $f_1, g_1, g_2, f_2$, 
$(ii)$ $f_1, f_2, g_2, g_1$, $(iii)$ $g_1, f_1, f_2, g_2$ and $(iv)$~$g_2, g_1, f_1, f_2$. For each case, when these $B$-segments are placed (see Figure~\ref{fig:cycle} and draw the four cases),
the positions of the $A$-origins $T, e_1, Q, e_2$ of the
8-cycle  are unique. Then, in cases $(ii)$, $(iii)$ and $(iv)$ it is possible to place $R$, but not $N$,
since in each case there exist two $B$-segments among $g_1, g_2, f_1, f_2$ that are separated from the others by $R$. The only
possible Stick representation for $\mathbb{H'}$, up to the symmetry , is the one in Figure~\ref{fig:gadgetO} (right) deduced from case $(i)$.

Furthermore, there are two possible positions for $P$ in the Stick representation with $f_1\prec f_2$ (immediately before or 
immediately after $N$). On its turn, $Y$ may be positioned everywhere next to $P$ or to $N$. 

There is no Stick representation with $f_2\prec f_1$. If, by contradiction,  we assume that $f_2\prec f_1$ can hold, then $\mathbb{H'}$ has a representation similar to the previous one,
where the indices $1$ and $2$ are switched. But there is no possible position for $P$ in that Stick representation, since this
is equivalent with placing in Figure~\ref{fig:gadgetO} (right) an $A$-segment $P'$ that would intersect only $g_1, f_1, g_2$, and this is not possible.

We deduce that the Stick representation with $f_1\prec f_2$ is the unique one, up to the positions of $P$ and $Y$.
\end{proof}

\subsection{The order gadget}

\begin{figure}[t!]
\centering
\hspace*{-3cm}\scalebox{0.6}{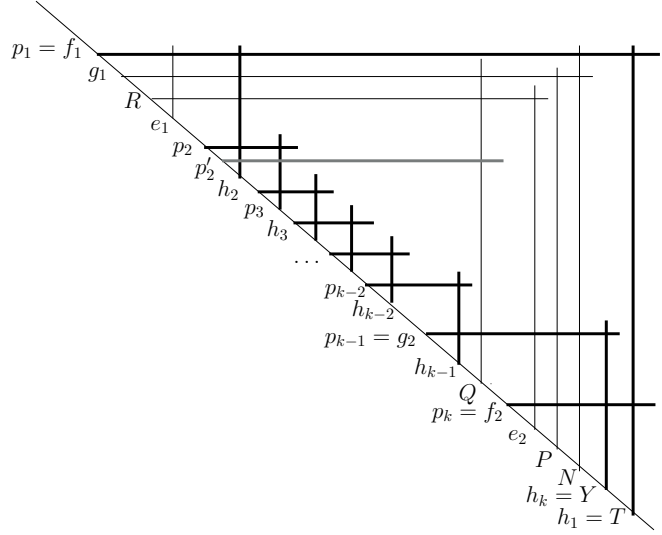}
\caption{\small Representation of a forced cycle. The thick segments represent the underlying cycle. The grey segment is $p'_2$. All the other segments form the order gadget. Only one possible position is shown for $P$ and $Y$.}
\label{fig:orderGrand}
\end{figure}

The order gadget is the handy gadget used in a particular context, aiming at ordering the variable and clause-transmission gadgets.
We start by showing that it imposes  only one, precise,  Stick representation for a  {\em forced cycle} $\mathbb{F}$, that is a graph 
obtained as described below. Figure~\ref{fig:orderGrand} illustrates the construction of a forced cycle directly using the Stick representation. 
This Stick representation is unique, as shown in Proposition~\ref{prop:order}.

\begin{itemize}
 \item Consider a cycle, called the {\em underlying cycle} of $\mathbb{F}$,  with vertices $h_1,p_1, h_2, p_2, \ldots, h_k,p_k$, where $h_i$ (respectively $p_i$) are $A$-vertices (respectively $B$-vertices).
 \item  Identify $p_1, p_{k-1}, p_k, h_k, h_1$ respectively with the vertices $f_1, g_2, f_2, Y$ and $T$ of a handy gadget (called in this
 context the {\em order gadget}), whose other vertices have the same notations as in Figure~\ref{fig:gadgetO}. 
 \item  Add a new vertex $p'_2$, as well as the edges $p'_2h_2, p'_2h_3, p'_2Q, h_2R, h_2g_1$. 
\end{itemize}

\bprop
Up to minor variants concerning the positions of $P$ and $Y$, a forced cycle $\mathbb{F}$ has a unique Stick representation (shown in Figure~\ref{fig:orderGrand}).   
\label{prop:order}
\eprop

\begin{proof}
By Remark \ref{rem:cycle} and given that the order of $f_1$ and $f_2$ is forced by the order gadget, the only possible orders from left to right
of the $B$-origins representing the vertices in the underlying cycle  are $p_1(=f_1), p_2, \ldots, p_k(=f_2)$ or $p_i, p_{i-1}, \ldots, p_2,$ $p_1(=f_1),$ $p_k(=f_2), p_{k-1}, \ldots, p_{i+1}$,
for some $i$ with $1\leq i \leq k-1$ (see also Figure~\ref{fig:cycle}). 
The latter case is not possible. To see this, note that the origin $p'_2$ is outside the cage $\lcage p_1, h_1\rcage$, 
since $p'_2$ intersects $h_2$ and $h_3$. Moreover, by Proposition~\ref{prop:stickRhandyG} and Figure~\ref{fig:gadgetO}, the segment $Q$ is
placed inside the cage $\lcage f_1,T\rcage$ (that is, $\lcage p_1, h_1\rcage$). Then $p'_2$ and $Q$ cannot intersect. 
 Thus the Stick representation in Figure~\ref{fig:orderGrand}, issued from the former case,
is the unique Stick representation possible for the graph $\mathbb{F}$, up to the positions of $P$ and $Y$ from the order gadget.
\end{proof}

\br
The reader may notice that the construction in Figure~\ref{fig:overview} allows to imagine the application of Proposition~\ref{prop:order}.  In our construction, the variable gadgets will 
have $B$-segments  playing the roles of $p_i$, $2\leq i\leq k-1$, and of $p'_2$. These $B$-segments together with the other order $A$-, $B$-segments from Figure~\ref{fig:overview}
and the order gadget will form a forced cycle, as in Proposition~\ref{prop:order}.
\er

By definition, the order gadget used in the construction of our graph is denoted by $\mathbb{O}$. Its intersections with other gadgets are progressively presented below.

\subsection{The clause gadgets}

Let $K=(x_i\vee x_j\vee x_k)$ be a clause, for which $i<j<k$ by convention. For simplicity reasons, let $x=x_i$, $y=x_j$, $z=x_k$. 
The three ways to satisfy the clause $K$ under the 1-in-3 SAT constraints are
TFF, FTF,  FFT, where each triplet (called a {\em truth triplet}) indicates the truth values of the literals, in the order $x, y, z$. 

The clause gadget $\mathbb{C}_K$ of $K$ is presented in Figure~\ref{fig:clauseGgraphOnly}. It contains, for each literal among $x, y,z$, a pair of $B$-origins 
whose order along the ground line encodes a {\em state} among T (true) and F (false).  
The literal $x$ (respectively $y,z$) is encoded by the pair of origins
$(b_1^1,b_1^0)$ (respectively $(a_1, \newc)$ and $(V,a_1)$) and has state T if and only if $b_1^1\prec b^0_1$
(respectively $a_1\prec \newc$, respectively $V\prec a_1$).  
The clause gadget admits four Stick representations (with several unimportant variants) that realize, for the triplet $x,y,z$ of literals, exactly three triplets of states, that we call {\em state triplets}. The state triplets are TFF,  FTF,  FFT, exactly as the
truth triplets admitted in a 1-in-3 SAT truth assignment for $K$.


We start with an intermediate result concerning the handy gadget $\mathbb{H}$ containing $H$ and $S$:

\bprop
The following properties hold for each Stick representation of the clause gadget $\mathbb{C}_K$:
\begin{enumerate}
\item[(a)] no proper segment of $\mathbb{C}_K$ (that is, different from $H$ and $S$), can have its origin between the first and the last origin in the Stick representation of 
$\mathbb{H}$.
\item[(b)] $b_1\prec b_2$. 
\end{enumerate}
\label{prop:propHandy}
\eprop

\begin{proof}
Consider a Stick representation of the clause gadget  $\mathbb{C}_K$. The handy gadget $\mathbb{H}$ forces $H$ to be positioned before $S$ (see Figure~\ref{fig:gadgetO}, with $e_1=H$, $N=S$).

Considering (a), note first that $b_2\prec e_1(=H)$ since $b_2$ and $H$ are adjacent. Moreover, $b_2\prec g_1$, 
otherwise $b_2$ -- which intersects $H$ -- would intersect $Q$ before intersecting $S$, a contradiction.   
Assume first, by contradiction, that $e_1\prec J\prec T$, for some proper segment $J\neq b_2$ of $\mathbb{C}_K$.  Then
$J$ is confined inside one of the cages $\lcage R,e_2\rcage$, $\lcage f_2,T\rcage$ that cover the interval delimited by $e_1$ and $T$. 
Then each path in $\mathbb{C}_K$ joining the two vertices $b_2$ and $J$ has a vertex whose corresponding segment intersects $r, e_2, f_2$ or $T$, and this is not the case since no vertex in $\mathbb{C}_K$
except $H$ and $S$ has this property. Thus no proper origin $J\neq b_2$ of $\mathbb{C}_K$ belongs to the interval defined by the
origins $e_1$ and $T$ of $\mathbb{H}$. Moreover, if some proper origin $L\neq S,H$ of $\mathbb{C}_K$ (here, $L=b_2$ is possible)
satisfies $f_1\prec L\prec R$ then all the proper origins of $\mathbb{C}_K$ satisfy the same relation. Indeed, from (i) no proper segment of
$\mathbb{C}_K$ intersects $f_1$ or $T$,  and (ii) no proper origin belongs to the interval defined by $e_1$ and $T$ (shown above), we deduce that
every segment that intersects $S$(=$N$) must also intersect $H$(=$e_1$), which is not true. Property (a) is proved.

\begin{figure}[t!]
\centering
\vspace*{-1cm}
\hspace*{-3cm}
\scalebox{0.5}{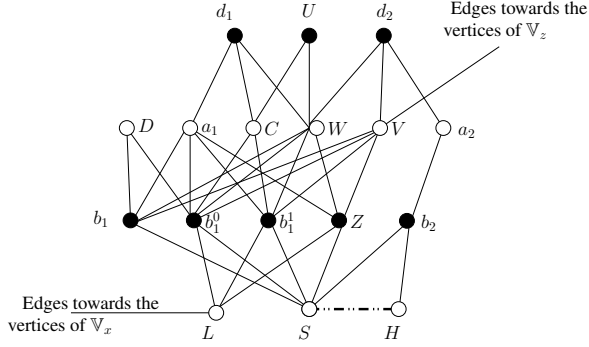}
\caption{\small Clause gadget $\mathbb{C}_K$. The dashed line between $H$ and $S$ means that the vertices $H$ and $S$ belong to a handy gadget (see Figure~\ref{fig:gadgetO}) where they are identified with the
local vertices $e_1$ and $N$, respectively. 
The vertex $L$ (respectively $V$) has neighbors in $\mathbb{V}_x$ (respectively $\mathbb{V}_z$), the variable gadget of $x$ (of $z$).}
\label{fig:clauseGgraphOnly}
\end{figure}

We also prove (b) by contradiction. If $b_2\prec b_1$, then $b_2\prec b_1\prec H\prec S$ by (a). Then, since $b_1, b_2$ intersect $H$, the segment $H$ cannot intersect $b_2$ unless it intersects $b_1$, a contradiction.  
\end{proof}
\bigskip

{\bf Conventions.} In the remainder of the paper, because of Proposition~\ref{prop:propHandy},  the expressions {\em immediately before $H$} and {\em immediately after $S$} have to be understood 
respectively as {\em immediately before $f_1$ from $\mathbb{H}$} and {\em immediately after $T$ from $\mathbb{H}$}. We also implicitly refer to  Proposition~\ref{prop:propHandy} when
we forbid to position any proper origin of $\mathbb{C}_K$ between $H$ and $S$.

%
%

\begin{figure}[t!]
\centering
\vspace*{-1cm}
\hspace*{-3cm}
\hspace*{1.5cm}\scalebox{0.57}{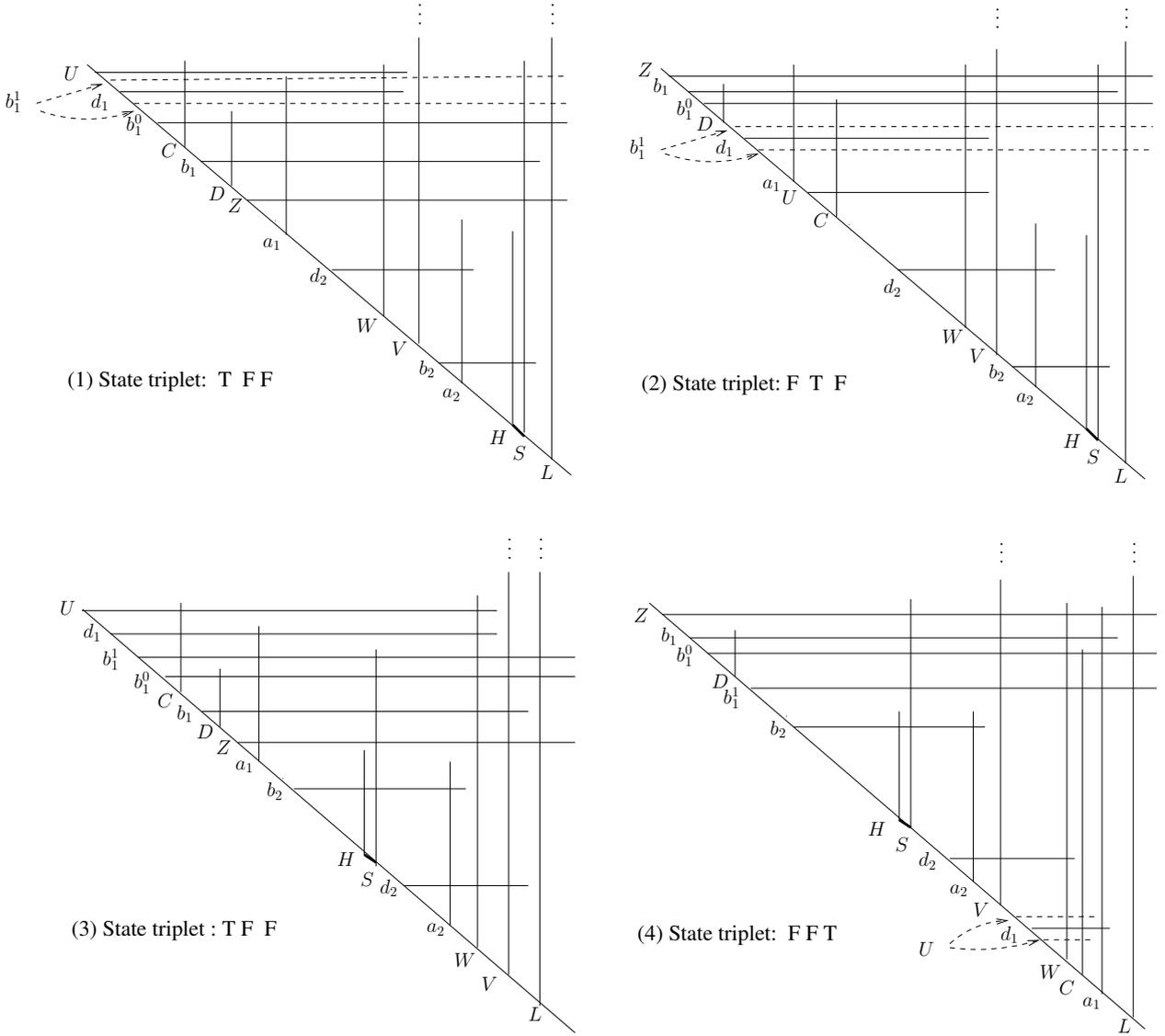}
\caption{\small The four Stick representations (and their minor variants) of the clause gadget $\mathbb{C}_K$.  The handy gadget
containing $H$ and $S$ is drawn as a thick segment along the ground line in the Stick representations.
The vertices $L,V$ have neighbors respectively in $\mathbb{V}_x$ and $\mathbb{V}_z$, the variable gadgets of $x$ and respectively $z$.
Each state triplet indicates the states of the literals $x,y,z$ in this order. The dashed segments show different possible positions for the same origin.}
\label{fig:clauseG}
\end{figure}

\bprop
In each standard Stick representation of $\mathbb{G}(X,\mathcal{K})$,
the clause gadget $\mathbb{C}_K$ has exactly four Stick representations (up to minor variants), given in Figure~\ref{fig:clauseG} together with the state triplets they encode.
\label{prop:clauseStickrepr}
\eprop

\begin{proof}
The possible Stick representations of $\mathbb{C}_K$ are computed by a fastidious case-by-case study. We assume that the segments of  $\mathbb{C}_K$ 
are successively drawn in an order which depends on the configuration at study.  
Each time a segment is added, all its intersections with the {\em already drawn} segments must be drawn. When several  positions are 
available for the origin of a segment, all of them are kept until a new segment (if any) allows to eliminate some of them.
We do not need to draw the handy gadget, we only consider its impacts, according to Proposition~\ref{prop:propHandy}. Four cases occur 
once the segments $b_1, b_2, S$ and $H$ are drawn, with $b_1\prec b_2\prec H\prec S$:
\medskip

{\it Case 1. $a_1, a_2\prec S$.} Then $b_1\prec a_1\prec b_2\prec a_2\prec H$.

There are  two possible positions  for $b_1^0$, either immediately before or immediately after $b_1$. As $d_2a_2\in E(\mathbb{C}_K)$ but $d_2a_1\not\in E(\mathbb{C}_K)$,
we have $a_1\prec d_2\prec a_2$, and with $Wd_2, Wb_1\in E(\mathbb{C}_K)$ but $Wb_2\not\in E(\mathbb{C}_K)$, we deduce that $a_1\prec d_2\prec W\prec b_2$.
We now consider the following subcases: 

\begin{itemize} 
 \item  Case 1.1. $b_1^0\prec b_1$. Then $d_1$ has three possible positions, before $b_1^0$, between $b_1^0$ and $b_1$ or after $b_1$. As
$\newc$ intersects $b_1^0$ and $d_1$ but not $b_1$, the position of $d_1$ after $b_1$ is not valid since it allows no possible place for $\newc$. Then there are two possible positions for $\newc$:
either immediately after both $d_1$ and $b_1^0$, regardless to the position of $d_1$, or immediately after $S$, in the case where 
$d_1\prec b_1^0$ (in this case $d_1$ also precedes the tip of $S$). The latter case is not possible, since then the segment $V$ -- which intersects $d_2$ -- necessarily intersects $d_1$ before intersecting $\mathbb{V}_z$, that is placed before the clause gadget in each standard Stick
representation of $\mathbb{G}(X,\mathcal{K})$, a contradiction. Then $\newc$
is positioned immediately after both $d_1$ and $b_1^0$, just before $b_1$. In this configuration,  $U$ -- which intersects $\newc$ and $W$ but
not $a_1$ -- is necessarily placed before both origins $d_1$ and $b_1^0$. Then $Z$ is positioned either immediately before or immediately
after $b_1$. Now, $\newa$ -- which does not intersect $Z$ -- has two possible positions: either $\newa$ is placed between $b_1$ and $Z$
(when $b_1\prec Z$ and $d_1\prec b_1^0$), or $\newa$ is placed after $S$. The latter case is impossible again. Indeed, $L$ cannot be situated before $S$,
since $L$ must intersect the variable gadget $\mathcal{V}_x$ of $x$ (which is situated before the clause gadget in each standard Stick representation), 
but $L$ intersects none of  $U, d_2$ and $b_2$. Therefore $L$ must be placed between $S$ and $\newa$, because the tip of $Z$ is situated before $\newa$.
But then $L$ cannot intersect $b_1^0$ unless it intersects $b_1$, a contradiction. We deduce that $b_1\prec \newa\prec Z$ and $d_1\prec b_1^0$. 
Then $b_1^1$, which does not intersect $D$,  has two possible places, next to $d_1$.
It remains to position $V$, whose unique possible place satisfies $W\prec V\prec b_2$, and $L$ for which only the position immediately after $S$ is possible.
We obtain the Stick representation in Figure~\ref{fig:clauseG}(1). With $b_1^1, b_1^0$ in this order we deduce that the state of the 
literal $x$ of the clause $K=(x\vee y\vee z)$ is fixed to T by this Stick representation. 
The orders of $a_1,\newc$ and of $V, a_1$ imply that the states of $y$ and $z$ are both~F.

\hspace*{-1cm}

\item Case 1.2. $b_1\prec b_1^0$. Then initially $d_1$ has, again, three possible positions. The position $d_1\prec b_1$ implies that  $S\prec \newc$ and, once more, 
this position is not valid since $V$ would be impossible to place. Then $d_1$ has two possible positions (immediately before or after $b_1^0$) and $\newc$ is placed after $d_1$ and $b_1^0$, and has two possible positions
(before or after $a_1$). However, the position of $U$ implies that $a_1\prec \newc$ is the only possible place for $\newc$, since we must have 
$a_1\prec U\prec \newc$. Now, $\newa$ can be placed either between $b_1^0$ and $d_1$ (implying that only the position of $d_1$ after 
$b_1^0$ is valid) or after $S$. In the latter case, as $b_1^1\newc\in E(\mathbb{C}_K)$ we deduce that $b_1\prec b_1^1$, thus the tip of $b_1^1$ precedes $\newa$.
Then $L$ -- regardless to
where it is situated -- cannot intersect $b_1^1$, $b_1^0$ and $\mathbb{V}_x$ unless
it intersects $b_1$, a contradiction. Thus $\newa$ is necessarily positioned between $b_1^0$ and $d_1$. Then $b_1^1$ must be placed
after $\newa$, either immediately before or immediately after $d_1$. There are unique positions for $Z, L$ and for $V$,
as shown in Figure~\ref{fig:clauseG}(2), which describes the Stick representation we just found.
With $b_1^0, b_1^1$ in this order we deduce that the state of the literal $x$ of the clause $K=(x\vee y\vee z)$ is fixed to F by this Stick representation. 
The orders of $a_1,\newc$ and of $V, a_1$ imply that the state of $y$ is T and the state of $z$ is F.
\end{itemize}

\medskip

{\it Case 2. $a_1\prec S\prec a_2$.} Then $b_1\prec a_1\prec b_2$.

We first show that $W$ cannot be placed before $S$. In this case $d_2$, that intersects $W$ and $a_2$, is necessarily positioned before $b_1$ and before the tip of $S$, otherwise $d_2$ would intersect
$S$ before $a_2$. Then $d_2\prec b_1^1$ since $b_1^1S\in E(\mathbb{C}_K)$ and the tip of $b_1^1$, which does not intersect $a_2$, lies between $S$ and $a_2$. Consequently, each vertical segment that intersects $b_1^1$ and reaches the variable gadgets (which are placed before the clause gadget since the Stick representation is standard)
must intersect $d_2$, and this contradicts the definition of $L$, which is one of these vertical segments but does not intersect $d_2$. 

Thus $S\prec W$ and $W$ must be situated after $a_2$ since $Wb_2\not\in E(\mathbb{C}_K)$. The origin $b_1^0$ (two possible positions, as before) is easily placed.  
The origin $d_2$ has a unique possible position (between $S$ and $a_2$). Furthermore,
$d_1$ intersects $a_1$ and $W$, and its only possible position is before $b_1$ and $b_1^0$. 

Concerning $\newc$, we have two cases. If $W\prec \newc$, then with $d_1\prec b_1\prec d_2\prec \newc$ we deduce that $V$ is between $d_2$ and $\newc$. Consequently, $V$ cannot intersect $\mathbb{V}_x$ unless 
it intersects $d_1$, a contradiction. So there is one convenient position for $\newc$, namely between $b_1^0$ and $b_1$ (implying that only the position $b_1^0\prec b_1$ is possible for $b_1^0$). Now, $U$ can be placed only before $d_1$ (since $Ua_1\not\in E(\mathbb{C}_K)$) and we consider the cases concerning $\newa$. 

\begin{itemize}
 \item If $\newa$ is immediately before or immediately after $a_1$, then $b_1^1$ -- which does not intersect $D$ -- can be positioned only immediately before $b_1^0$.
Moreover, $Z$ and $L$ have unique positions, and imply that $\newa\prec Z\prec  a_1$. Then $V$ must be placed between $W$ and $L$. We have obtained the  Stick representation in Figure~\ref{fig:clauseG}(3). 
With $b_1^1, b_1^0$ in this order we deduce that state of the literal $x$ is fixed to T by this Stick representation, whereas 
the orders of $a_1,\newc$ and of $V, a_1$ imply that both the states of $y$ and $z$ are F.

\item If $\newa$ is after $W$, then $Z$ is still necessarily placed after $\newc$, thus its tip lies between $W$ and $\newa$. But $L$
cannot be situated before $W$, since it would intersect $d_1$ and $U$ before it intersects $\mathbb{V}_x$. And $L$ cannot
be situated between $W$ and $\newa$ since then it would intersect $b_1$ before it intersects $\mathbb{V}_x$, a contradiction.
Thus this case cannot appear.
\end{itemize}
\medskip

{\it Case 3. $S\prec a_1,a_2$.} 

Then $S\prec a_2\prec a_1$, otherwise $a_1$ intersects $b_2$, a contradiction. Again, $b_1^0$ (two possible positions each) is easily placed. Then $W$ has a unique possible position, between $a_2$ and $a_1$, after the
tip of $b_2$.  As $d_2$ intersects $a_2$ and $W$ but not $S$,
we deduce that $d_2$ must be positioned immediately before $a_2$. We have two possibilities with respect to $d_1$, which may be placed either between $a_2$ and $W$
or before $b_1$ and $b_1^0$.

\begin{itemize}
 \item If $a_2\prec d_1\prec W$, then $\newc$ either immediately precedes or immediately follows $a_1$. In the latter case, $U$
 is necessarily placed before $b_1$ and $b_1^0$, and also before $Z$ since $Z$ intersects $S$ and  the tip of $S$ lies after $U$. But then the
 tip of $Z$ lies between $a_1$ and $\newc$, and $L$ -- which intersects $Z$ -- cannot intersect the variable gadget of $x$
 unless it intersects $U$, a contradiction. Thus $\newc$  is positioned between $W$ and $a_1$. From $\newc b_1\not\in E(\mathbb{C}_K)$ and $\newc b_1^0\in E(\mathbb{C}_K)$
 we deduce that $b_1^0$ must be placed after $b_1$. Then $U$ can be positioned only next to $d_1$, before or after it. We have again two cases, this time with respect to $\newa$, which must intersect $b_1$
 and $b_1^0$. If $\newa$ is situated immediately after $b_1^0$, then $b_1^1$ can be placed only after $\newa$. Concerning $Z, L$ and $V$,
 they have unique places as indicated in Figure~\ref{fig:clauseG}(4). The state triplet for 
 the literals $x, y,z$ is thus FFT. If $\newa$ is positioned after $a_1$, then $b_1^1$ -- whose possible places are immediately before
 or immediately after $b_1^0$ -- has its tip between $a_1$ and $\newa$. Thus $L$, which must intersect $b_1^1$, cannot intersect
 $\mathbb{V}_x$ unless it intersects $b_1$, a contradiction.
 \item If $d_1\prec b_1, b_1^0$, then  $V$, which intersects $d_2$, must intersect $d_1$ before it reaches
 the variable gadget $\mathbb{V}_z$, a contradiction.
\end{itemize}

{\it Case 4. $a_2\prec S\prec a_1$.} Then $b_2\prec a_2\prec H$.

The only valid position for $W$, so that $d_2$ may be placed too, is immediately before $b_2$, and thus $b_1\prec d_2\prec W\prec b_2$. Then $d_1\prec b_1$
and, as before, $V$ cannot be placed so as to ensure all its intersections, and a non-intersection with $d_1$.
\end{proof}

\paragraph{Connections with the other gadgets} 
The $B$-segments $b_1^0$ and $b_1^1$ are also the two order $B$-segments outgoing from the clause gadget of $K$ to the right. 
See Figure \ref{fig:overview}.  Moreover, for each clause $K$, there is an $A$-order segment associated with it, which intersects the $B$-order segments outgoing from the clause $K$,
as well as those outgoing from the previous clause (or intersects the last variable gadget, if $K=K_1$). See Figure~\ref{fig:overview}. 

Consequently, except when $K=K_m$,  $b_1^0$ and $b_1^1$ intersect the order $A$-segment immediately following the clause and the order $A$-segment following the next clause.
When $K=K_m$,  $b_1^0$ and $b_1^1$ intersect by definition the order $A$-segment associated to the clause $K$, and another $A$-segment (the rightmost order $A$-segment in  Figure~\ref{fig:overview}).
This segment corresponds to the  $A$-segment $h_{k-1}$ in Figure~\ref{fig:orderGrand}, and -- again by definition -- intersects in addition the segment $g_2$ of the order gadget $\mathbb{O}$ used 
to build $\mathbb{G}(X,\mathcal{K})$.

Finally, $b_1^0$ and $b_1^1$ intersect the transmission gadgets of all the literals in the clause following $K$, when $K\neq K_m$. More precisely,  $b_1^0$ and $b_1^1$ intersect 
all the $A$-segments from these transmission gadgets that go from their clause gadget towards a variable gadget (shorter $A$-segments exist in some transmission gadgets, but their tips do not go farther than the origins of the
corresponding clause gadget and are not concerned here). As it will be seen later, the segments $L$ and $V$ from $\mathbb{C}_K$ will be included in the transmission gadgets of $x$ and $z$,
and are thus concerned by these intersections (more details are given later).

\subsection {The transmission gadgets}

The Stick representation of the transmission gadget for a literal $t\in\{x, y, z\}$ is roughly a set of vertical segments that have 
to transmit the state of the literal $t$ in the clause $K=(x\vee y\vee z)$ towards the variable
gadget $\mathbb{V}_t$. There are two main difficulties here: (a) the transmission gadget for a fixed literal must intersect exactly the same horizontal
segments in all the four Stick representations of the clause $K$, and (b) it must correctly transmit to the variable gadget 
the state of the literal. 

\begin{figure}[t!]
\centering
\hspace*{-3cm}\scalebox{0.6}{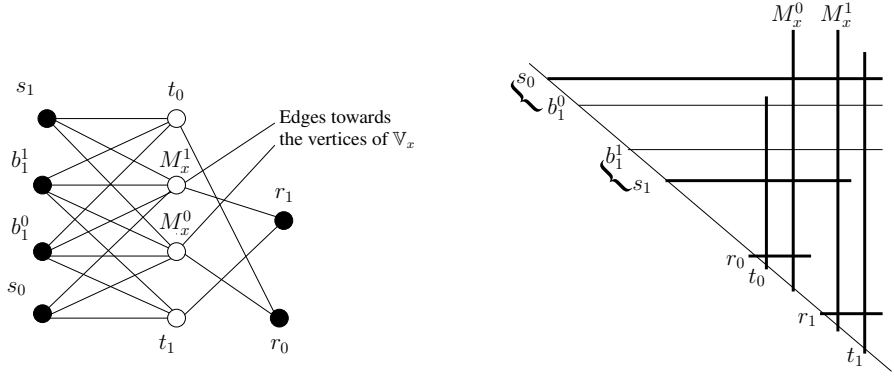}
\vspace*{-2cm}
\caption{\small Left: Transmission gadget $\mathbb{T}_x$ of the literal $x$, and its intersections with $b_1^0$ and $b_1^1$. By definition, $N_{\mathbb{C}_K}(s_h)=N_{\mathbb{C}_K}(b_1^h)$ holds for
$h=0,1$. Right: its unique Stick representation (up to minor variants that keep the same order for the pairs of origins $(s_0,s_1), (r_0,r_1), (M_x^0,M_x^1)$) when $b_1^0\prec b_1^1$. 
The ''$\{$'' marks indicate that the two origins are inseparable, because of the condition  $N_{\mathbb{C}_K}(s_h)=N_{\mathbb{C}_K}(b_1^h)$, $h=0,1$.}
\label{fig:ordretransmissionxy}
\end{figure}

The transmission gadget $\mathbb{T}_x$ of $x$ and one of its Stick representations are drawn in Figure~\ref{fig:ordretransmissionxy}. Intuitively, $M_x^0, M_x^1$ are two $A$-segments 
whose origins must have the same order as $b_1^0, b_1^1$, and which transmit this order to the variable gadget of $x$. The segments  $s_0, s_1, t_0, t_1$ are (almost) duplicates of $b_1^0, b_1^1, M_x^0, M_x^1$ respectively, 
and are inseparable from their originals along the ground line, when $\mathbb{T}_x$ is represented {\em in the context of} $\mathbb{C}_K$ (or, more generally, of 
$\mathbb{G}(X,\mathcal{K})$). That means they behave identically in terms of positions, due to the common neighborhood of $s_h$ and $b_1^h$ ($h=1,2$)
in $\mathbb{C}_K$, and to the common neighbor $r_h$ of $M_1^h$ and $t_h$ ($h=1,2$). Unlike their originals, the duplicates may be as short as we need and this allows us to obtain  
consistent orders between $b_1^0, b_1^1$ (that force $s_0$ and $s_1$) on the one hand, and $M_x^0, M_x^1$ (forced by $t_0, t_1$, which are
on their turn forced by $s_0, s_1$) on the other hand.

\begin{figure}[t!]
\centering
\hspace*{-1cm}\scalebox{0.52}{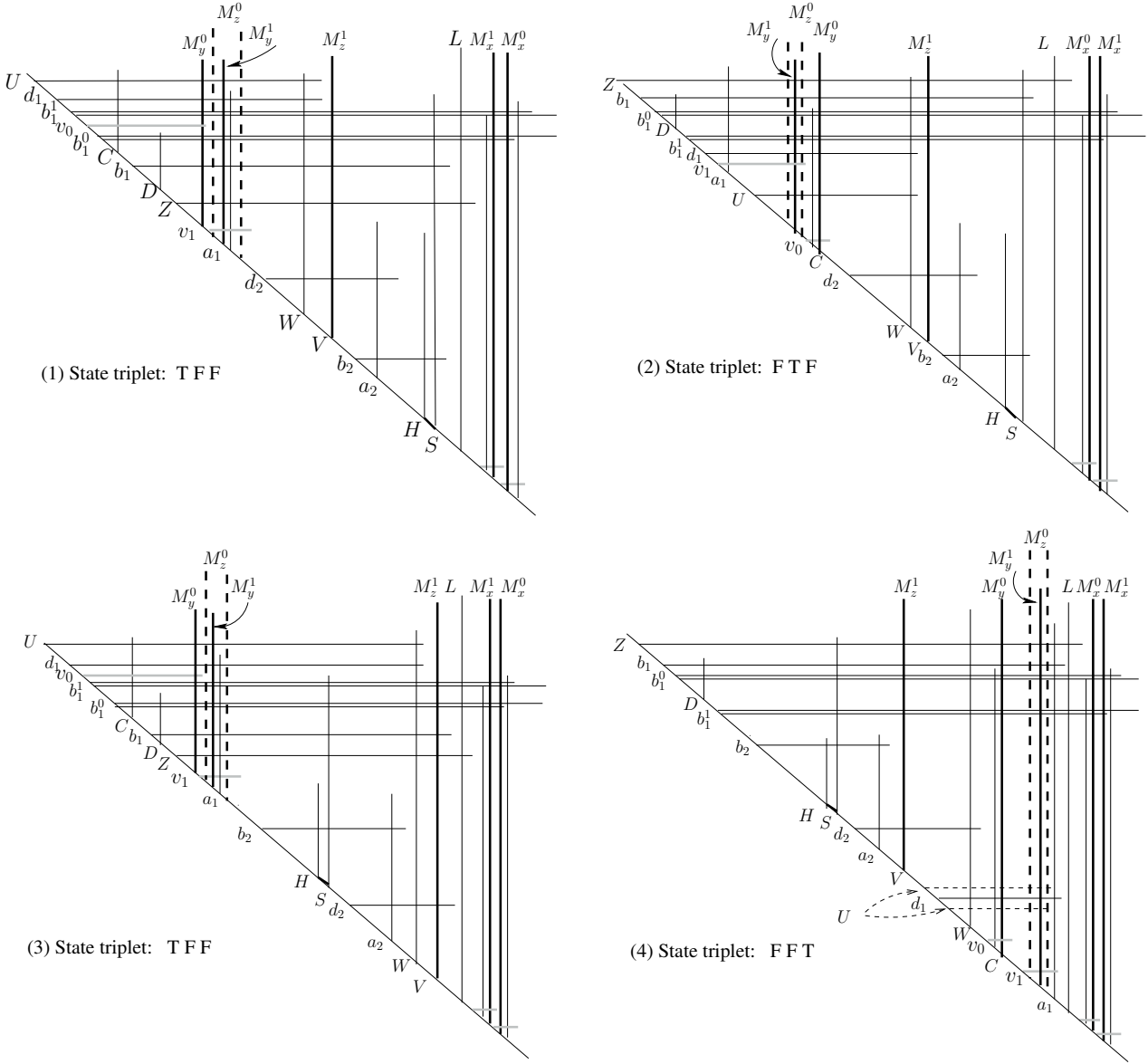}
\caption{\small Stick representations of the clause-transmission gadgets.  For simplification, in cases (1) and (2) we
chose without loss of generality one of the two possible positions for $b_1^1$. The duplicates of $b_1^0, b_1^1, M_x^0, M_x^1$ are drawn next to their originals, but their origins
are not named on the figure. All the $A$-segments that must join the variable literals - $L$ included - have their names on the top of the Stick
representation. Thick segments: the supports (except $L$). Grey segments:  $r_0, r_1, v_0, v_1$.  Dashed segments: the two valid positions for $M_z^0$ (before or after $M_y^1$).}
\label{fig:TransmissionG}
\end{figure}

The transmission gadget of $x$ is necessarily positioned in the Stick representations of the clause gadget $\mathbb{C}_K$ (given in Figure~\ref{fig:clauseG}) as indicated in Figure
\ref{fig:TransmissionG}, since $M_x^0, M_x^1, t_0, t_1$ do not have other neighbors but those indicated in Figure~\ref{fig:ordretransmissionxy}.
Note that whereas the origins $r_0,r_1$
may have several valid positions in the  Stick representation of $\mathbb{T}_x$  when $\mathbb{T}_x$ is considered separately, they have
unique convenient places (as in Figure~\ref{fig:TransmissionG}) when $\mathbb{T}_x$ is seen in the context of the clause gadget $\mathbb{C}_K$. The segments $r_0,r_1$ are drawn
in grey in Figure~\ref{fig:TransmissionG}, but their names are omitted.

\br
Given the position of $L$ with respect to the transmission gadget of $x$, and the fact that $L$ has neighbors in $\mathbb{V}_x$, we assume from now on that $L$ belongs to $\mathbb{T}_x$.
The intersection of $L$ with the other gadgets will be defined in this context.
\label{rem:Laussi}
\er

The transmission gadget $\mathbb{T}_y$ of $y$ is made of two $A$-segments $M_y^0$ and $M_y^1$, that are respectively duplicates of $\newc$ and $a_1$ with additional neighbors. 
More precisely $M_y^0$ is adjacent with $b_1^0, b_1^1$, their duplicates, $d_1,U$ (as $\newc$) and moreover with $Z, b_1$, whereas  $M_y^1$ is adjacent with $b_1,b_1^0, b_1^1$, their duplicates, $d_1,Z$ (as $a_1$)
and moreover with $U$. In order to keep the origins $\newc$ and $M_y^0$ (respectively $a_1$ and $M_y^1$) close to each other, both vertices are adjacent with a new vertex
$v_0$ (respectively $v_1$). The segments $v_0, v_1$ are also drawn in grey in  Figure~\ref{fig:TransmissionG}. They force $M_y^0, M_y^1$ to be placed as in the figure, since no other places are possible.

The transmission gadget $\mathbb{T}_z$ of $z$ is made of the $A$-segments $M_z^0$ and $M_z^1$ defined as follows: $M_z^0$ is a copy
of $M_y^1$ (they both represent the origin $a_1$) and is thus adjacent with 
$v_1, Z, U, d_1, b_1, b_1^0, b_1^1$ and the duplicates of $b_1^0, b_1^1$; whereas $M_z^1$ is equal to $V$. 
Then, in the Stick representations,  $M_z^0$ and $M_y^1$ are positioned in the neighborhood of $a_1$ in an arbitrary order, as indicated in Figure
\ref{fig:TransmissionG}, whereas $M_z^1$ is already placed as $V$.

\bprop
In each standard Stick representation of the graph $\mathbb{G}(X,\mathcal{K})$, the transmission gadgets satisfy: 

$M_x^1\prec M_x^0$ if and only if $b_1^1\prec b_1^0$,

$M_y^1\prec M_y^0$ if and only if $a_1\prec \newc$,  and 

$M_z^1\prec M_z^0$ if and only if $V\prec a_1$.
\label{prop:correcttransmission}
\eprop

\begin{proof}
 The affirmation is immediate for $M_y^0, M_y^1, M_z^1, M_z^2$ since the positions indicated in Figure~\ref{fig:TransmissionG} for these segments define
 a fixed order between pairs, in each case.

 The transmission gadget $\mathbb{T}_x$ has a unique place in each Stick representation of the clause gadget, but $M_x^0$ and $M_x^1$ have the same neighborhoods in $\mathbb{C}_K$
 and they could possibly be switched without switching $b_1^0$ and $b_1^1$. We show that the duplicates force the origins $M_x^0$ and $M_x^1$ to have the same order as the origins $b_1^0$ and $b_1^1$. 
 Assume by contradiction  that we could have $b_1^0\prec b_1^1$ and $M_x^1\prec M_x^0$.  See Figure~\ref{fig:ordretransmissionxy} and consider a Stick representation of $\mathbb{T}_x$ and $\mathbb{C}_K$, 
 as in Figure~\ref{fig:TransmissionG}. Then with $b_1^0\prec b_1^1$, we are either in case (2) or in case (4), but $M_x^0$ and $M_x^1$ are switched
 with respect to the illustrations in Figure~\ref{fig:TransmissionG}.  
 In each of these cases, we  position $s_0$ and $s_1$ next to  $b_1^0$ and respectively $b_1^1$ (since $N_{\mathbb{C}_K}(s_h)=N_{\mathbb{C}_K}(b_1^h)$, $h=0,1$). Note that
 in case (2), $d_1$ could separate $b_1^1$ and $s_1$, since the two possible positions for $b_1^1$ in Figure \ref{fig:clauseG} case (2) are
 also possible for $s_1$, but this is not important. The vertex $t_1$ is adjacent with $b_1^0, b_1^1$ and $s_0$ but not to $s_1$, so $M_x^0\prec t_1$ 
 or $b_1^1\prec t_1\prec s_1$. In both cases, $r_1$ must be placed before $s_1$, and its tip is after $M_x^1$. But then $r_1$ intersects $V$
 (which must intersect $\mathbb{V}_z$), a contradiction.
 The case where $b_1^1\prec b_1^0$ and $M_x^0\prec M_x^1$ is similar since  the construction is symmetric.
%
\end{proof}

According to Proposition~\ref{prop:correcttransmission}, the $A$-segments $M_x^0, M_x^1$, $M_y^0, M_y^1$, $M_z^0$ and $M_z^1$
record the states of the literals $x, y, z$, in this order,  in the Stick representation of $\mathbb{C}_K$. These segments, together with $L$ (see Remark~\ref{rem:Laussi}),
are respectively called the {\em supports} of $x, y, z$. 

\paragraph{Connections with the other gadgets} 
In order to ensure the consistency with the standard Stick representation of $\mathbb{G}(X,\mathcal{K})$, assume now that $K\neq K_1$. We impose by definition that, for each literal $u\in\{x,y,z\}$ from $K$, 
the supports of $u$ intersect the $B$-segments $b_1^0, b_1^1$ outgoing from the clause gadget $\mathbb{C}_{K'}$, where $K'$ is the clause immediately preceding $K$ in the standard order of the clauses
(and thus of the clause gadgets).

\subsection{The variable gadgets}

\begin{figure}[t!]
\centering
\hspace*{-1cm}\scalebox{0.5}{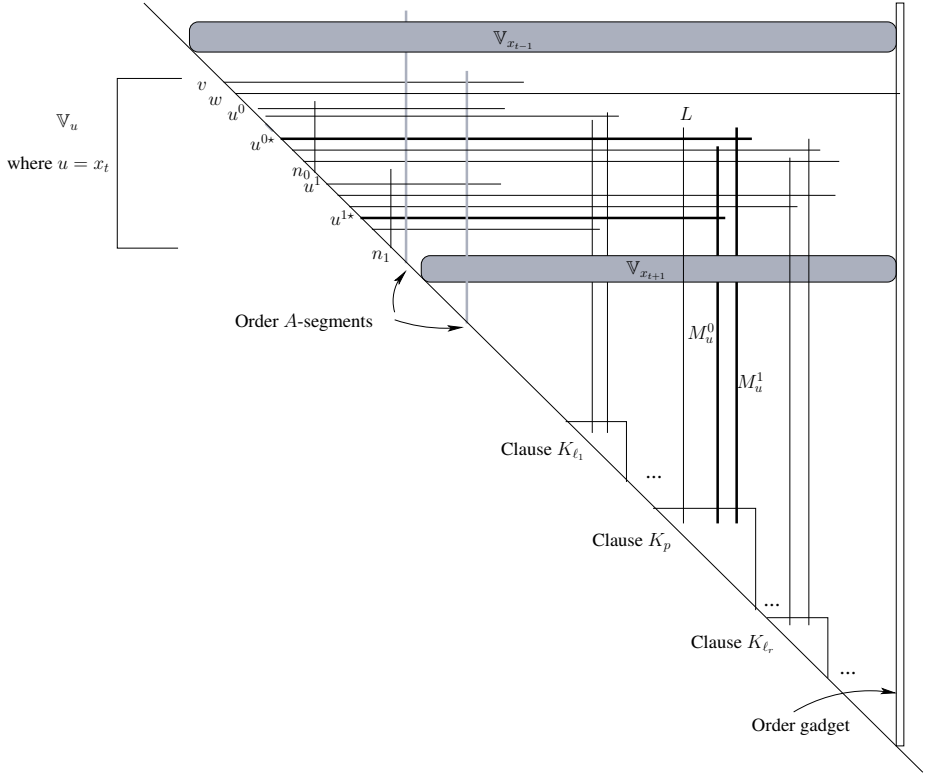}
\caption{\small  Variable gadget $\mathbb{V}_u$ for $u=x_t$ and its intersections with the transmission gadgets outgoing from the clause $K_p$. The thick segments indicate the connection 
with the transmission gadget $\mathbb{T}_u$, when $u$ is any literal in the clause $K_p=(x\vee y\vee z)$. The segment $L$ is present only when $u=x$ in the clause $K_p$.}
\label{fig:VarG}
\end{figure}

Let $u$ be a variable among $x_1, x_2, \ldots, x_n$ (say $u=x_t$), and let $K_{\ell_1}, {K}_{\ell_2}, \ldots, {K}_{\ell_r}$ be the clauses in which $u$ occurs as a literal. 
The variable gadget $\mathbb{V}_u$ of $u$ is given in Figure~\ref{fig:VarG}, using directly its Stick representation. It has two types of vertices.

\paragraph{Vertices that force the order in  the standard Stick representation: $v$ and $w$} 
The neighbors of $v$ are the two order $A$-segments that intersect the variable gadget
according to Figure~\ref{fig:overview}.
The vertex $w$ has one more neighbor, namely $Q$ in the order gadget $\mathbb{O}$. The pair $(v,w)$ of segments in {\em each} variable
gadget is similar to the pair $(p_2,p'_2)$ in Figure~\ref{fig:orderGrand}. The order $A$-segments that intersect $v,w$ are
similar with $h_2, h_3$ in Figure \ref{fig:orderGrand}.

The use of the other vertices in $\mathbb{V}_u$ is presented later. By now, it is sufficient to notice (see Figure~\ref{fig:VarG}) that  they form two groups, that (a) contain respectively
$u^0$ and $u^1$, (b) are adjacent respectively with $n_0$ and with $n_1$, and (3) are both adjacent with the two order $A$-segments that intersect $v$ and $w$.

We are now ready to prove that:

\bprop
Each Stick representation of $\mathbb{G}(X,\mathcal{K})$ whose variable gadgets, clause gadgets, order segments and order gadget are defined as above is a standard Stick representation.
\label{prop:reprStickordonnee}
\eprop

\begin{proof}
 The variable gadgets and the order $A$-segments have been defined such that each order $A$-segment intersects the $B$-segments $v$ of two successive variable gadgets according to the increasing order of the variables. 
 On their turn, the clause gadgets have been defined such that each
 order $A$-segment intersects the order $B$-segments $b_1^0$ outgoing from two consecutive clauses according to the increasing order of the clauses, except for the first
 and last clause which are particular but for which the intersections with order $A$-segments are defined similarly.
 Thus the order $A$-segments, the order $B$-segments from the clause gadgets, the $B$-segments $v$ from the variable gadgets as well as the $B$-segment $w$ from the first variable gadget
 induce together with the order gadget a forced cycle. By Proposition~\ref{prop:order}, the only possible Stick representation for this forced cycle is the one in
 Figure~\ref{fig:orderGrand}, where the segments $h_i$ ($i=2, \ldots, k-1$) are the order $A$-segments, the segments $p_j$ $(j=2, \ldots, k-2)$
 are the order $B$-segments (the segments $v$ from the variable gadgets and the segments $b_1^0$ from the clauses gadgets) and 
 the unique vertex $w$ we have chosen is the vertex $p'_2$.
 
 It remains to show that all the origins of a variable gadget (respectively of a clause-transmission gadget) must be placed consecutively in an interval containing the
 origin $v$ belonging to the variable gadget (respectively the origin $b_1^0$ belonging to the clause gadget). For each variable or clause-transmission gadget $\mathbb{D}$ and 
 in any Stick representation of the graph, let the interval $I_{\mathbb{D}}$ be defined along the ground line by the first and the last origin 
in the gadget. Then $I_{\mathbb{D}}$ is called the  {\em underlying interval} of the gadget $\mathbb{D}$ in the Stick representation. Note that:
 
 \begin{enumerate}
\item[(a)]  There exists a set of $h$ cages $\lcage B_i, A_i\rcage$ with $1\leq i\leq h\leq 2$ 
covering $I_{\mathbb{D}}$, such that the origin $B_1$ is the leftmost origin of $I_{\mathbb{D}}$, and $A_h$ is either (i) the rightmost origin in $I_{\mathbb{D}}$ or (ii) the $A$-origin
immediately following the rightmost origin in $I_{\mathbb{D}}$.
The clause-transmission gadgets satisfy  $h=2$ and (i), whereas the variable gadgets either satisfy $h=1$ and (ii), or satisfy  $h=2$ and (i).
For instance, in the Stick representation in case (1) of a clause-transmission gadget (see Figure~\ref{fig:TransmissionG}), the cages defined by $\lcage U,W\rcage$ and by the duplicates 
 of $b_1^1$ and of $M_x^0$ cover the underlying interval. For the variable gadget represented in Figure~\ref{fig:VarG}, the underlying interval is covered by the cage made of $v$ and the 
order $A$-segment immediately following $n_1$, which immediately follows $n_1$ (that belongs to the gadget). Thus $h=1$ and (ii) holds. However, the order $A$-segment could precede $n_1$ 
in a slightly different Stick representation, in which case we would have $h=2$ and property (i). 
\item[(b)] No origin of an order $A$-segment may be included in the underlying interval of a clause-transmission gadget.  Indeed, Figure~\ref{fig:TransmissionG} shows that no position exists inside the interval
for an $A$-segment that would intersect only $b_1^0$ and $b_1^1$ (because of the duplicates of $b_1^0$ and $b_1^1$, and the segments $r_0,r_1$ from $\mathbb{T}_x$). 
\item[(c)] The segment $Q$ from the order gadget $\mathbb{O}$ cannot be included in a cage $C$ from the set of cages covering the underlying interval $I_{\mathbb{D}}$ of a variable or clause-transmission gadget. 
If, by contradiction, this was the case, then the
cage $C$ would contain the entire order gadget. This is due to the observation that in the sequence $Q, R, e_1, f_1, T$ from the order gadget each segment intersects the next one, see Figure~\ref{fig:gadgetO} and  Figure~\ref{fig:orderGrand},  
but none of them intersects any cage from any other gadget, thus none of them may have one endpoint in $C$ and the other endpoint out of $C$. Consequently, the forced cycle defined above -- which has all its
origins between $f_1$ and $T$ -- is also included in the same cage $C$, and thus $\mathbb{D}$ is not a clause-transmission gadget because of (b).
Then $\mathbb{D}$ is a variable gadget $\mathbb{V}_s$, one of whose segments precedes $f_1$ and one of whose $A$-segments $n_t$, $t\in\{0,1\}$,
follows $T$. But then $u^t$ and all its copies follow $T$ (otherwise they should intersect $h_1$ in order to intersect $n_t$, a contradiction). And thus the order $A$-segments intersecting $u^t$ should follow $h_1$,
which contradicts Proposition~\ref{prop:order}.
\end{enumerate}
\medskip

%
%

We need to show that the set $\mathcal{I}$ defined as 
\medskip

$\mathcal{I}=\{I_{\mathbb{D}}\, |\, \mathbb{D} \text{ is a variable gadget or a clause-transmission gadget}\}$ 
\medskip

\noindent contains only disjoint intervals. In this case,
any Stick representation would be a standard Stick representation, since each order $B$-segment from the forced cycle with be replaced on the ground line with the underlying interval of the gadget it belongs to.

By contradiction, assume that the set $\mathcal{I}$ contains two non-disjoint underlying intervals $I_{\mathbb{D}}$ and $I_{\mathbb{D'}}$.
Then $I_{\mathbb{D}}$ and $I_{\mathbb{D'}}$ cannot strictly overlap, otherwise two cages from those covering $I_{\mathbb{D}}$ and $I_{\mathbb{D'}}$ according to (a) would necessarily intersect, resulting into at least one wrong segment 
intersection. And if $I_{\mathbb{D}}$ and $I_{\mathbb{D'}}$ are contained in each other, for instance $I_{\mathbb{D}}\subsetneq I_{\mathbb{D'}}$, we have that:
\begin{itemize}
\item $\mathbb{D}$ is a clause-transmission gadget. Indeed, if $\mathbb{D}$ was a variable gadget, the  segment $w$ of $\mathbb{D}$ would be inside a cage $C$ from the set of cages covering $I_{\mathbb{D'}}$, whereas the segment $Q$ from
the order gadget $\mathbb{O}$ would be outside $C$ by (c). Then one cannot have $wQ\in E(\mathbb{G}(X,\mathcal{K}))$ as required.
\item $\mathbb{D'}$ is a variable gadget. To see this, let $\mathbb{C}_K$ be the clause gadget in the clause-transmission gadget $\mathbb{D}$.
The origin $b_1^0$ from $\mathbb{C}_{K}$ belongs to $I_{\mathbb{D}}$, and thus to $I_{\mathbb{D'}}$, but cannot intersect any cage from 
$\mathbb{D'}$. Consequently, the order $A$-segment $h_d$ (where $n+2\leq d\leq n+m+2$ is an integer)  associated with the clause $K$ (which intersects $b_1^0$ by definition) is inside $I_{\mathbb{D'}}$. Then (b) implies
that $\mathbb{D'}$ is not a clause-transmission gadget.
\end{itemize}

Consequently, assume that $\mathbb{D'}$ is the variable gadget $\mathbb{V}_{u}$, and let  $n_s$, $s\in\{0,1\}$, be the rightmost origin among $n_0$ and $n_1$ from 
$\mathbb{V}_{u}$. Then $n_s$ is the right endpoint of $I_{\mathbb{D'}}$, and thus the order $A$-segment $h_d$ seen above (with $n+2\leq d\leq n+m+2$) satisfies $h_d\prec n_s$ (since, as shown above, the order $A$-segment  $h_d$ is inside $I_{\mathbb{D'}}$). Moreover, the origin of the segment $u^s$ from $\mathbb{V}_{u}$ precedes the origin $h_e$ $(2\leq e\leq n+1$) of any of the two order $A$-segments that intersect
$\mathbb{V}_u$, since $u^sh_e\in E(\mathbb{G}(X,\mathcal{K}))$. Thus
$u^s\prec h_e\prec h_d\prec n_s$. But then, according to the representation of $\mathbb{O}$ (Figure \ref{fig:orderGrand} with $k=n+m+3$), $n_s$ intersects one of the segments $p_d, p_{d+1}, \ldots, p_{k}$ ({\em i.e.} either one of the $B$-segments $b_1^0$ issued from the clause gadgets with the same clause numbers, or $g_2$, $f_2$ from the handy gadget) before it intersects $u^s$. But none of these segments is allowed to intersect $n_s$,
thus we have a contradiction.

This concludes the proof.
\end{proof}

\paragraph{Vertices that force the variable consistency}

In the variable gadget, $u^0$ and $u^1$ are the two $B$-segments that record the (global) state of the variable
$u$ in the following way. See Figure~\ref{fig:VarG}. Each of $u^0$ and $u^1$ has $r$  {\em copies} of different lengths, with one pair of copies 
devoted to each clause containing $u$. The copies of $u^s$ ($s=0,1$) are always consecutive due to the $A$-segment $n_s$ that
intersects them all (and no other segment).  Let us call $u^{0\star}, u^{1\star}$ the copies devoted to a fixed clause $K_p$, where $p=\ell_i$
for some $i\in\{1, 2, \ldots, r\}$. With the aim of ensuring that  $M_u^1$ is before $M_u^0$ in a Stick
representation if and only if $u^{1\star}$ is before $u^{0\star}$, we use the construction in 
Figure~\ref{fig:VarG} (thick segments) that is similar to the one we used for the duplicates in Figure~\ref{fig:ordretransmissionxy}: 
we require the intersection between $M_u^0$ and  $u^{1\star}$ but not $u^{0\star}$, and the intersection between $M_u^1$
and $u^{0\star}$ but not $u^{1\star}$. Moreover, both $M_u^0$ and $M_u^1$ intersect all the copies of $u^0$ and $u^1$ that are
longer than $u^{0\star}$ and $u^{1\star}$, {\em i.e.}, that are devoted to clauses $K_{\ell_s}$ with $\ell_s>p$. Furthermore,
if $K_p=(x\vee y\vee z)$ and $u=x$, then the segment $L$ from $\mathbb{T}_u$ is adjacent to all the copies of $u^0$ and $u^1$ that are
longer or equal to $u^{0\star}$ and $u^{1\star}$, {\em i.e.}, that are devoted to clauses $K_{\ell_s}$ with $\ell_s\geq p$. See
Figure \ref{fig:VarG}.

\begin{figure}[t!]
\centering
\hspace*{-1cm}\scalebox{0.30}{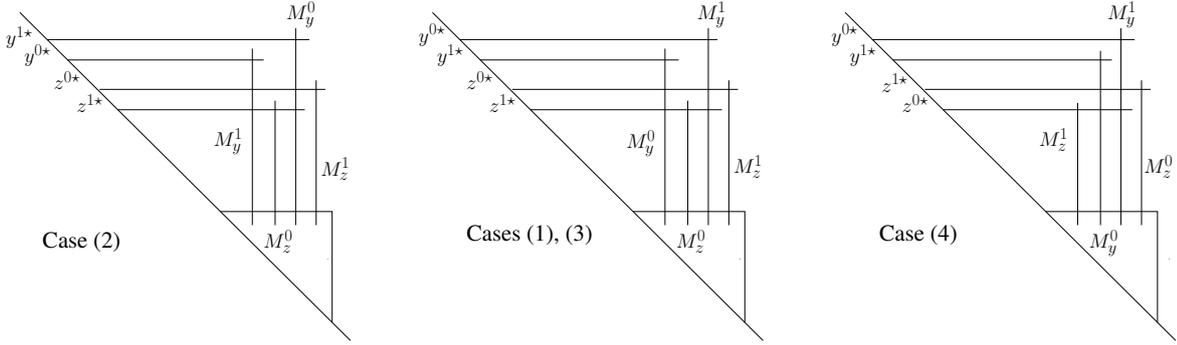}
\caption{\small Specific intersections between  $\mathbb{T}_y$ and $\mathbb{V}_z$ from the clause $K=(x\vee y\vee z)$, and the resulting Stick representations in each of the four cases in
Figure~\ref{fig:TransmissionG}. In cases (1), (2) and (3), the two possible positions of $M_z^0$ accept these intersections (just move $M_z^0$ on the opposite side of $M_y^1$). In case (4), only the
position of $M_z^0$ after $M_y^1$ is valid, and the other one is eliminated at this stage.}
\label{fig:VariableGyz}
\end{figure}

This construction is valid for $u\in\{x,y,z\}$. However, the supports of $y$ and $z$ are intertwined (see Figure \ref{fig:TransmissionG}), and this imposes intersections between the supports
 $M_y^1$ and $M_y^0$ and the segments $z^{1\star}$ and $z^{0\star}$, as shown in Figure~\ref{fig:VariableGyz}. In each case the intersections are the same,
 and the figure shows that all the three configurations admit a Stick representation of these intersections.

\bprop
Let $u\in \{x_1, x_2, \ldots, x_n\}$. Then:
\begin{itemize}
 \item[(a)] For each clause $K_p=(x\vee y\vee z)$ containing $u$, if $u^{0\star}, u^{1\star}, M_u^0$ and $M_u^1$ are the notations used above, then the left to right order of the origins $u^{0\star}, u^{1\star}$
 is the same as the left to right order of $M_u^0$ and $M_u^1$, in each Stick representation of these four segments.
 \item[(b)] For each variable $u$, the origin $u^0$ and all its copies are positioned before $u^1$ and all its copies, or vice versa.
 \label{prop:varG}
\end{itemize}

\eprop

\begin{proof}
In all cases, the intersection of $M_u^1$ with $u^{0\star}$ but not with $u^{1\star}$, and of $M_u^0$ witn $u^{1\star}$ but not with $u^{0\star}$ implies that if exactly one pair of origins among
$(u^{0\star}, u^{1\star})$ and $(M_u^0,M_u^1)$ is switched then a wrong intersection appears. The second affirmation is implied by the existence of the common neighbor $n_0$ (respectively $n_1$) of $u^0$
(respectively $u^1$) and all of its copies.
\end{proof}

\paragraph{Connections with the other gadgets} According to the standard Stick representations, we impose by definition that each support of $u$ from the clause $K_p$ intersects, for each $h>t$ (recall that $u=x_t$), all the
$B$-segments of $\mathbb{V}_{x_h}$ devoted to clauses $K_j$ with $j>p$. This is equivalent to requiring that, for $u^{0\star}$ and $u^{1\star}$ devoted to $K_p$, both the segments  
$u^{0\star}$ and $u^{1\star}$, but also the segment $w$ of $\mathbb{V}_u$, intersect all the supports of the literals $x_q$ with $q<t$ belonging to clauses $K_j$ with $j<p$.

Concerning the two A-order segments shown in Figure \ref{fig:VarG}, we required that they both intersect all the $B$-segments in $\mathbb{V}_u$, and we used the leftmost of them in Proposition~\ref{prop:reprStickordonnee}. One exception
exists to these rules: the second (rightmost) order $A$-segment for the last variable $x_m$ is the order $A$-segment associated with the first clause $K_1$. Consequently, it must intersect, by definition, only the
$B$-segments from $\mathbb{V}_{x_m}$ that are devoted to clauses $K_p$ with $p>1$ ({\em i.e.}, which are long enough to intersect the order $A$-segment rising after the clause gadget of $K_1$). 

%
 
The construction of the graph $\mathbb{G}(X,\mathcal{K})$ associated with the set $X$ of variables and the set $\mathcal{K}$ of clauses is now complete.

\section{Main theorem}\label{sect:mainth}

\bthm
 {\sc StickRec} is NP-complete.
\ethm

\begin{proof}
Note first that {\sc StickRec} belongs to NP, since -- unlike SEG graphs -- in Stick graph representations the segments lie in a fixed number of directions and this is an essential feature for NP membership \cite{kratochvil1994intersection}.

Next, we show that {\sc Monotone 1-in-3SAT}  with input $X$ and $\mathcal{K}$ has a positive answer if and only if the graph $\mathbb{G}(X,\mathcal{K})$ is a Stick graph.

Consider $X$ and $\mathcal{K}$, as well as the graph $G(X,\mathcal{K})$ defined above. By Proposition~\ref{prop:reprStickordonnee}, each Stick representation
 of  $\mathbb{G}(X,\mathcal{K})$, if such a representation exists, is a standard Stick representation. The incidental intersections between clause gadgets, variable gadgets, 
 transmission gadgets and order segments have been defined according to the standard Stick representation, as indicated in Remark \ref{rem:intersectionvartrans} and further applied
 for each gadget (see the ''Connections with other gadgets'' paragraphs). They are therefore satisfied in each standard Stick representation. The existence of a Stick representation
 for the graph $G(X,\mathcal{K})$ is thus equivalent to the existence of a Stick representation for the gadgets such that the literal states
 output by the clause gadgets using the transmission gadgets are
 consistent, which means that for each variable the states of the literals equal to it in different clauses are the same.
 
We put the truth values (T or F) for a variable $u$ in $X$ in bijection with the Stick representations of the corresponding variable gadget $\mathbb{V}_u$ as follows:
 the T (respectively F) value for $u$ is represented by a variable gadget with $u^1\prec u^0$ (respectively $u^0\prec u^1$). By Proposition~\ref{prop:varG}(b), the latter
 affirmation holds if and only if each pair of copies $u^{1\star}$ and $u^{0\star}$ of $u^1$ and $u^0$ follows the same precedence relation. By  Proposition~\ref{prop:varG}(a) and Proposition~\ref{prop:correcttransmission},
 the order of $u^{1\star}$ and $u^{0\star}$ along the ground line correctly records the state of the literal $u$ in any clause gadget containing $u$. 
 Thus, the value affected to  $u$ is T (respectively F) if and only if each occurrence of the literal $u$ in a clause gadget returns the state T (respectively F) to the variable gadget. 
 
 Now, assume that {\sc Monotone 1-in-3SAT} has a positive answer for the instance $(X,\mathcal{K})$. Then, each clause $K$ has a truth triplet among FFT, FTF, TFF, implying -- according to
 the above bijection -- a state triplet identical to the truth triplet for the clause gadget $\mathbb{C}_K$. Since each state triplet among  FFT, FTF, TFF is Stick representable 
 (Proposition~\ref{prop:clauseStickrepr}), and all the states of the literals are consistent (because, by hypothesis, we have a truth assignment for the variables in $X$), we deduce that
 $G(X,\mathcal{K})$ has a Stick representation. 
 
 Conversely, if $G(X,\mathcal{K})$ has a Stick representation, then each clause gadget realizes one of the state triplets FFT, FTF, TFF  (Proposition~\ref{prop:clauseStickrepr}) and, by the equivalence we
 establish above, each variable is affected a value T or F that corresponds to all the occurrences of this variable in the clause gadgets. Then, for each clause $K$ whose
 clause gadget has a state triple FFT, FTF, or TFF, its truth triplet is the same, and the conclusion follows.   
\end{proof}

\section{Recognizing BipHook and Max Point-Tolerance graphs}\label{sect:biph}

Max point-tolerance graphs (also known as {\em hook graphs}, {\em heterozygosity graphs} or {\em p-BOX(1)} graphs) are the superclass of BipHook graphs defined as the intersection graphs of hooks whose centers lie on a line with slope -1. They were defined several times (see \cite{catanzaro2017max,halldorsson2011clark,hixon2013hook,thraves2015p}) and
have several characterizations, but recognizing them is an open problem.

We thus consider the following questions.
\bigskip

\noindent{\sc BipHook Graph Recognition (BipHookRec)}

\noindent {\bf Input:} A bipartite graph $\Gamma$.

\noindent{\bf Output:} Does $\Gamma$ admit a BipHook representation?
\bigskip

\noindent{\sc Max Point-Tolerance Graph Recognition (MPT-Rec)}

\noindent {\bf Input:} A graph $\Gamma$.

\noindent{\bf Output:} Is $\Gamma$ a max point-tolerance graph?
\bigskip

Both problems belong to NP, since the intersection of hooks reduces to the intersection of segments in two directions (only the interpretation of
the intersections is different with respect to Stick graphs).

\bthm
{\sc BipHookRec} is NP-complete.
\ethm

\begin{proof} 
The reduction is done from {\sc StickRec}. Consider an instance $G=(A\cup B,E)$ of {\sc StickRec}, and assume without loss of generality that $G$ is a connected graph. Build an instance $\Gamma$ of {\sc BipHookRec} as
follows:
\medskip

\begin{itemize}
 \item for each vertex $u$ of $G$, add to $\Gamma$ an induced 4-cycle $F_u$ with  $V(F_u)=\{x_u, t_u, y_u, z_u\}$ and  $E(F_u)=\{x_ut_u, t_uy_u, y_uz_u, z_ux_u\}$.
 \item for each edge $uv$ of $G$, let $x_u$ and $y_u$ be adjacent to $x_v$ and $y_v$.
\end{itemize}
\medskip

\noindent Note that $\Gamma$ is a bipartite graph with the following bipartition: 
\medskip

$W_1=\{x_a\, |\,a\in A\}\cup  \{y_a\, |\,a\in A\}\cup \{t_b\, |\,b\in B\}\cup \{z_b\, |\,b\in B\}$

$W_2=\{x_b\, |\,b\in B\}\cup  \{y_b\, |\,b\in B\}\cup \{t_a\, |\,a\in A\}\cup \{z_a\, |\,a\in A\}$
\medskip

We show that $G$ is a Stick graph if and only if $\Gamma$ is a BipHook graph. 
\medskip

For the forward direction, let us start with a Stick representation of $G$. Then we build a BipHook representation of $\Gamma$ by replacing each  $B$-segment $b$ (each $A$-segment $a$, respectively) in the Stick representation of $G$ with the BipHook representation of $F_b$ ($F_a$, respectively)  presented
in Figure \ref{fig:StickHook} top (bottom, respectively). Then each edge $ab$ from $G$ is represented as in Figure \ref{fig:StickHook}, and the
resulting representation is a BipHook representation of $\Gamma$.

\begin{figure}[t!]
\centering
\hspace*{-1cm}\scalebox{0.30}{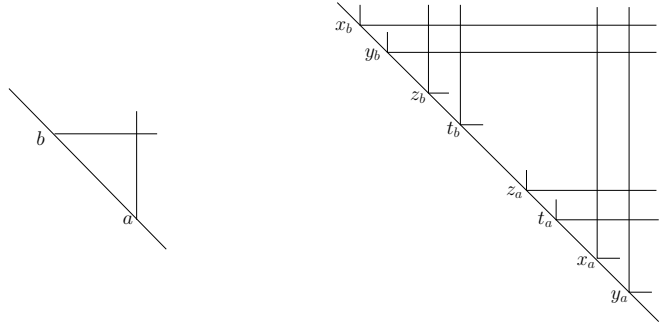}
\caption{\small Transforming a Stick representation of $G$ into a BipHook representation of $\Gamma$.}
\label{fig:StickHook}
\end{figure}

We now show the backward direction. Consider a BipHook representation of $\Gamma$ and let $F_u$ be the induced 4-cycle associated in  $\Gamma$ with an arbitrary vertex $u$ from $G$. Simplify
the notation by removing the index $u$ for $F_u$ and its vertices, which become therefore $F$ and $x, t, y, z$. We have the following property:
\medskip

(P1) Up to the symmetry of $x,y$ and that of $t,z$, there are exactly four distinct BipHook representations for $F$, shown in Figure \ref{fig:4repr}.
\medskip

Indeed, once $x$ and $y$ are represented by the hooks $H_x$ and $H_y$, at most one of the hooks $H_t$ and $H_z$ has its center between 
$x$ and $y$ (otherwise, $H_t$ and $H_z$ would intersect, a contradiction). When $z$ (for instance) satisfies $x\prec z\prec y$, we obtain
the BipHook representations 1 and 2, whereas when none of $z$ and $t$ is placed between $x$ and $y$ we obtain the representations 3 and 4. 

The
representations 1 and 3 (2 and 4, respectively) are called of {\em type $B$} ({\em type $A$}, respectively), since the horizontal (vertical, respectively) segments of the hooks $H_x$ and $H_y$ can be as long as possible. Intuitively, $A$-type BipHook representations of $F$
mimic the $A$-segments in a Stick representation of $G$, and similary for $B$-type Hook representations. 

We now show that:
\medskip

(P2) $uv\in E(G)$ if and only if the BipHook representations of $F_u$ and $F_v$ are of different types. Moreover, there exists a
BipHook representation of $\Gamma$ such that the following property holds for each $uv\in E(G)$: 

\begin{tabular}{ll} 
 (Q)&\begin{minipage}[t]{14.5cm}
if $F_u$ is of type $B$ and
$F_v$ is of type $A$, then the intersections between $H_{x_u}, H_{y_u}$ (on one side) and $H_{x_v}, H_{y_v}$ (on the other side) hold on the
horizontal segments of $H_{x_u}, H_{y_u}$ and the vertical segments of $H_{x_v}, H_{y_v}$.\end{minipage}
\end{tabular}
\medskip

Let us first consider the case where the BipHook representation of $F_u$ is the representation 1 (of type $B$). Then $H_{x_v}$ and $H_{y_v}$, which do not intersect 
$H_{z_u}$ but intersect $H_{x_u}$ and $H_{y_u}$ necessarily satisfy  $y_u\prec x_v$ and $y_u\prec  y_v$. Consequently, the representation of $F_v$ cannot be of type $B$, since then either $H_{y_v}$ (when $x_v\prec y_v$) or $H_{x_v}$ (when $y_v\prec x_v$) does not intersect $H_{y_u}$, a contradiction. We deduce that the
representation of $F_v$ is of type $A$, and the intersections between hooks satisfy Property (Q). A similar reasoning holds when the BipHook representation of $F_u$ is the representation 2.
\begin{figure}[t!]
\centering
\hspace*{-1cm}\scalebox{0.40}{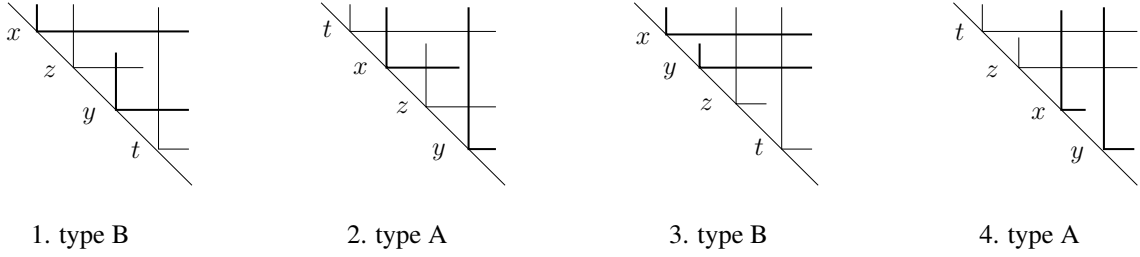}
\caption{\small The four possible representations of a 4-cycle $F$, and their types.}
\label{fig:4repr}
\end{figure}

In the case where the BipHook representation of $F_u$ is the representation 3 (of type $B$), the subcase where both $x_v$ and $y_v$ are located
after $y_u$ is similar to the previous case. We consider now the subcase where $x_v$ (without loss of generality)  satisfies $x_u\prec x_v\prec y_u$. Then $H_{y_v}$, which intersects both $H_{x_u}$ and $H_{y_u}$ but not $H_{x_v}$ satisfies $y_u\prec y_v$, with three possible
positions for the center $y_v$: between $y_u$ and $z_u$, between $z_u$ and $t_u$, or after $t_u$. Then the center $z_v$ of $H_{z_v}$ can be 
placed either before $x_u$ (regardless to the place of $y_v$), or between $x_u$ and $x_v$ (only when $y_u\prec y_v\prec z_u$). The same
constraints hold for the center $t_v$ of $H_{t_v}$. But then, in all the resulting configurations, the representation of $F_v$ is the representation 4, of type $A$. The intersections between hooks do not satisfy Property (Q) yet, because of the intersection between
$H_{x_v}$ and $H_{y_u}$. Notice that no center of a hook exists between $x_v$ and $y_u$. Indeed, since $G$ is connected, $\Gamma$ is connected too,
and then at least one of the hooks whose center lies between $x_v$ and $y_u$ should intersect at least one of the hooks $H_{x_v}$ and $H_{y_u}$.
But the neighbors of $H_{x_v}$ and $H_{y_u}$ are already placed in the BipHook representation, and their centers are not between $x_v$ and $y_u$.
We deduce that no center of a hook exists between $x_v$ and $y_u$, and then the center of $x_v$ can be moved immediately after $y_u$ so that
to satisfy Property (Q) for $uv$. A similar reasoning holds when the BipHook representation of $F_u$ is the representation 4.

We now finish our proof that if $\Gamma$ is a BipHook graph then $G$ is a Stick graph. Consider the BipHook representation of $\Gamma$ whose
existence is proved in (P2). And consider the subgraph $H$ of $\Gamma$ induced by $M\cup N$, where:
\medskip

$M=\{x_t\in V(\Gamma)\,|\, F_t\hbox{ is of type }A\}$

$N=\{x_w\in V(\Gamma)\,|\, F_w\hbox{ is of type }B\}$
\medskip

Then $H$ is isomorphic with $G$, since $M\cup N$ contains one vertex from each $F_u$ ($u\in V(G)$), and 
$x_ux_v\in E(H)$ if and only if $uv\in E(G)$, by the definition of $\Gamma$. Moreover, 
if $x_ux_v\in E(H)$, then by (P2) we have either $x_u\in M$ and $x_v\in N$ or vice-versa.
Given that $G$, and thus $H$, is connected, the bipartition of $G$, and thus of $H$, is unique,
implying that we have either $M=A$ and $N=B$, or $M=B$ and $N=A$. In the first case, the second part of  (P2) implies 
that the intersection between hooks always holds on the vertical segment for the vertices in $A$, and on
the horizontal segment for the vertices in $B$. The BipHook representation
of $H$ is easily transformed into a Stick representation of $G$ by removing the horizontal segments of the hooks
corresponding to vertices in $A$, and the vertical segments of the hooks corresponding to vertices in $B$.
In the second case, a similar approach results into a Stick representation where the vertices in $A$ are
represented with horizontal segments and those in $B$ with vertical segments. Reversing the order of the origins 
on the ground line and switching the directions of the segments (vertical versus horizontal) allows to obtain 
a Stick representation with vertical segments for the vertices in $A$ and horizontal segments for those in $B$. 
\end{proof}

{\sc BipHookRec} and {\sc MPT-Rec} are equivalent on bipartite graphs. We therefore deduce that:

\begin{cor}
{\sc MPT-Rec} is NP-complete.
\end{cor}

\section{Conclusion}\label{sect:conclusion}

We showed in this paper that recognizing Stick, BipHook and max point-tolerance graphs are NP-complete problems. Many other questions remain open. One of them concerns the complexity of the recognition problem for the immediate superclasses of
BipHook, namely 3-dimensional GIG, stabbable GIG and segment-ray graphs, all of which are also grid intersection graphs. We refer the reader to \cite{chaplick2018grid}
for the definition of these classes and their relations with other classes of intersection graphs. 
Another set of open problems consists in characterizing these classes by forbidden subgraphs. 
 \bibliographystyle{plain}
 \bibliography{Stick4Revision}
\end{document}